\documentclass[sigconf]{acmart} 
\usepackage{float}
\usepackage{subcaption}
\usepackage{caption}
\usepackage{listings}
\usepackage{xcolor}
\usepackage{multirow}
\usepackage{booktabs}
\usepackage{tabularx}
\usepackage{enumitem}
\definecolor{promptbg}{gray}{0.95}
\definecolor{promptframe}{gray}{0.5}
\lstdefinestyle{prompt}{
  basicstyle=\ttfamily\footnotesize,
  breaklines=true,
  breakindent=0pt,
  columns=fullflexible,
  frame=single,
  rulecolor=\color{promptframe},
  backgroundcolor=\color{promptbg},
  showstringspaces=false,
  upquote=true,
  aboveskip=6pt,
  belowskip=6pt,
  xleftmargin=4pt,
  xrightmargin=4pt,
  framexleftmargin=4pt,
  framexrightmargin=4pt,
  tabsize=2,
  keepspaces=true,
}

\usepackage{soul}
\usepackage{pifont} 
\usepackage{mdframed,xcolor}
\newmdenv[
  roundcorner=6pt,
  linecolor=black!20,
  linewidth=0.8pt,
  backgroundcolor=black!2,
  innertopmargin=6pt,
  innerbottommargin=6pt,
  innerleftmargin=8pt,
  innerrightmargin=8pt
]{catbox}

\hypersetup{
  pdftitle={Toward Scalable Audio Description Quality Control: A Workflow for Evaluating Human and VLM Raters},
  pdfauthor={},
  pdflang={en}
}

\AtBeginDocument{%
  }

\copyrightyear{2026}
\acmYear{2026}
\setcopyright{cc}
\setcctype{by}
\acmConference[ASSETS '26]{The 28th International ACM SIGACCESS Conference on Computers and Accessibility}{October 25--28, 2026}{Vila Nova de Gaia, Portugal}
\acmBooktitle{The 28th International ACM SIGACCESS Conference on Computers and Accessibility (ASSETS '26), October 25--28, 2026, Vila Nova de Gaia, Portugal}
\acmDOI{10.1145/3797867.3829038}
\acmISBN{979-8-4007-2521-0/2026/10}

\begin{document}
\title{Toward Scalable Audio Description Quality Control: A Workflow for Evaluating Human and VLM Raters}

\author{Lana Do}
\email{do.ng@northeastern.edu}
\affiliation{%
\department{Khoury College of Computer Science}
  \institution{Northeastern University}
  \city{San Jose}
  \state{California}
  \country{USA}
}
\author{Gio Jung}
\email{gjung1@sfsu.edu}
\affiliation{%
\department{Department of Computer Science}
  \institution{San Francisco State University}
  \city{San Francisco}
  \state{California}
  \country{USA}
}

\author{Andrew Taylor Scott}
\email{axs2561@mavs.uta.edu}
\affiliation{%
\department{Computer Science and Engineering}
  \institution{University of Texas at Arlington}
  \city{Arlington}
  \state{Texas}
  \country{USA}
}

\author{Juvenal Francisco Barajas}
\email{jbarajas8@mail.sfsu.edu}
\affiliation{%
\department{Department of Computer Science}
  \institution{San Francisco State University}
  \city{San Francisco}
  \state{California}
  \country{USA}
}

\author{Alexander Mario Blum}
\email{AMBlum@enrichyouracademics.com}
\affiliation{%
  \institution{Enrich Your Academics}
  \city{Emeryville}
  \state{California}
  \country{USA}
}

\author{Shasta Ihorn}
\email{sihorn@sfsu.edu}
\affiliation{%
\department{Department of Psychology}
  \institution{San Francisco State University}
  \city{San Francisco}
  \state{California}
  \country{USA}
}

\author{Vassilis Athitsos}
\email{athitsos@uta.edu}
\affiliation{%
\department{Computer Science and Engineering}
  \institution{University of Texas at Arlington}
  \city{Arlington}
  \state{Texas}
  \country{USA}
}

\author{Ilmi Yoon}
\email{i.yoon@northeastern.edu}
\affiliation{%
\department{Khoury College of Computer Science}
  \institution{Northeastern University}
  \city{San Jose}
  \state{California}
  \country{USA}
}

\renewcommand{\shortauthors}{Do et al.}

\begin{abstract}

Digital video is central to communication, education, and entertainment, but without audio description (AD), blind and low-vision users are excluded. While crowdsourced platforms and vision-language models (VLMs) expand AD production, quality is rarely checked systematically. Existing evaluations rely on n-gram overlap metrics and short-clip guidelines, leaving open the question of how to assess long-form AD quality at scale. To address this, we developed a methodological workflow using Item Response Theory to evaluate VLM and human rater proficiency against expert-established ground truth. Evaluations were based on a six-dimensional framework, grounded in professional guidelines and shaped by insights from our accessibility experts and blind consultants. Findings suggest that top-performing VLMs can approximate ground-truth ratings at levels comparable to human raters. However, qualitative analysis reveals that VLM reasoning is less reliable and actionable than that of human respondents. These insights underscore the potential of hybrid evaluation systems that leverage VLMs alongside human oversight, offering a path toward scalable AD quality control.
\end{abstract}

\begin{teaserfigure}
  \centering
  \includegraphics[width=\textwidth]{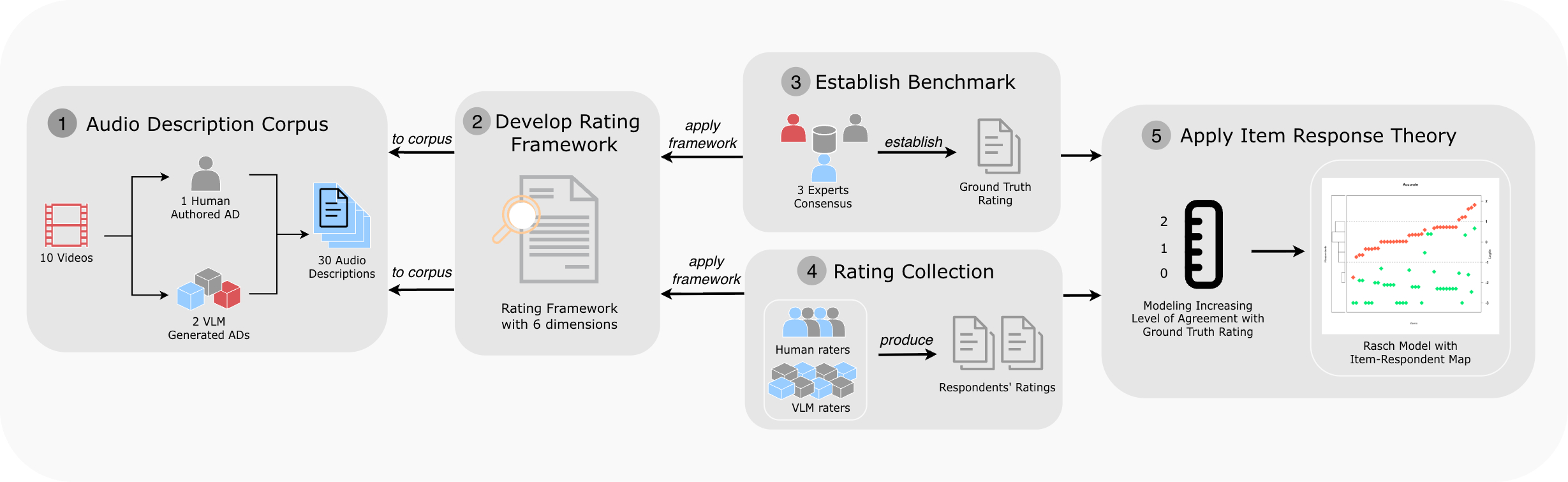}
  \vspace{-1em}
  \caption{Conceptual diagram of the five major stages in our evaluation workflow. From 10 videos, we collected 30 audio descriptions (1 human-authored and 2 VLM-generated per video) to form the AD corpus (\ding{172}). We developed a rating framework with six quality dimensions (\ding{173}). Three accessibility experts applied this framework to the corpus, establishing ground-truth ratings by consensus (\ding{174}); additional respondents, human and VLM raters, applied the same framework to the same corpus, producing the respondents' ratings (\ding{175}). We then applied Item Response Theory to model respondents' increasing levels of agreement with the ground truth, jointly estimating rater proficiency and item difficulty (\ding{176}).}
  \Description{The figure shows the conceptual diagram of the five major stages in the evaluation workflow. Starting from the left, Stage 1, titled "Audio Description Corpus," contains an icon of 10 videos pointing to icons of 1 human-authored and 2 VLM-generated audio descriptions, which in turn point to an icon of 30 audio descriptions. Stage 2, titled "Develop Rating Framework," contains an icon of a document labeled "Rating Framework with 6 dimensions," with an arrow labeled "to corpus" pointing from Stage 2 to Stage 1. Stage 3, titled "Establish Benchmark," contains an icon of 3 experts in consensus with an arrow labeled "apply framework" pointing toward Stage 2, and an arrow labeled "establish" pointing toward the Ground Truth Rating icon. Stage 4, titled "Rating Collection," mirrors Stage 3: icons of human raters and VLM raters with an arrow labeled "apply framework" pointing toward Stage 2, and an arrow labeled "produce" pointing toward the Respondents' Ratings icon. Stages 3 and 4 both point toward Stage 5, titled "Apply Item Response Theory," which contains a ruler icon representing the modeling of increasing levels of agreement with the ground truth rating, pointing toward an icon of a Rasch model with an item–respondent map.}
  \label{fig:experimental_workflow}
\end{teaserfigure}



\maketitle

\section{Introduction}
Audio description (AD) narrates key visual information in videos, enabling blind and low-vision (BLV) audiences to access media that would otherwise be inaccessible. Professional AD enhances comprehension and inclusion, but producing it is resource-intensive, limiting coverage given the scale of digital content available today \cite{PackerAnGuidelines}. To expand access, volunteer-driven authoring tools have emerged to lower barriers to AD creation through streamlined interfaces and collaborative editing \cite{YouDescribeYouDescribe.Https://www.youdescribe.org/, Pavel2020Rescribe:Descriptions, BranjeLiveDescribe:Description}. More recently, automated approaches have aimed to further extend coverage by leveraging large language models (LLMs) and vision–language models (VLMs) to generate scene-aware descriptions \cite{Wang2021TowardVideos, VanDaele2024MakingSummaries, Chu2024LLM-AD:System}. While these approaches expand coverage, their quality is often inconsistent, which can negatively impact the viewing experiences of BLV users. Studies on AD authoring tools such as Rescribe show that novices, when not aided by guidance, make timing and placement errors that reduce the usability of their descriptions \cite{Pavel2020Rescribe:Descriptions}. Similarly, AI-based systems have been shown to misidentify characters or hallucinate details \cite{li2025vidhalluc}, produce verbose or repetitive narration that confuses rather than supports viewers \cite{Wang2021TowardVideos}, and achieve high benchmark scores while misaligning with human judgments of AD quality \cite{Hu2024FIOVA}. These shortcomings highlight that expanding AD coverage through volunteer or AI contributions must be accompanied by systematic quality checks.

Existing AD quality assessments face two main limitations. First, they typically focus on isolated short clips, which fails to reflect real-world practice, whereas both volunteer and AI-generated AD now span uninterrupted full-length videos \cite{Liu2025VALOR:Dataset, Zhou2017TowardsVideos, Wang2019VATEX:Research}. Second, they rely on user studies with BLV participants, crowdsourced workers, or professional experts \cite{Li2025VideoA11y:Description}. While valuable, such studies are resource-intensive, and ratings from non-experts can be unreliable \cite{Webb2024TooTurk}. Traditional metrics like accuracy scores or unweighted averages add another layer to this unreliability, collapsing subjective judgments into a single number while ignoring variation in task difficulty and rater ability.

Our study addresses these gaps by adapting established principles from education and psychology to a novel context in accessibility research and Human-Computer Interaction (HCI). Utilizing Item Response Theory (IRT), we jointly model item difficulty and rater proficiency, rather than flattening ratings into aggregate scores. Developed by an interdisciplinary team of experts in assistive technology, measurement, test construction, and quality control, this workflow yields more reliable results and reveals nuanced insights that simple accuracy metrics overlook. Operationalizing this workflow required a measurement framework suited to full-length audio descriptions. To guide this development and subsequent application, we ask two research questions:
\begin{enumerate}
    \item \textbf{RQ1.} What constitutes AD quality in practice, and how can professional guidelines and expert knowledge be structured into a systematic rating framework?
    \item \textbf{RQ2.} Can VLMs function as reliable respondents within this workflow? Specifically, how do their ratings compare to human raters in aligning with expert ground truth, and what does this reveal about their potential for scalable, automated AD evaluation?
\end{enumerate}

To address \textbf{RQ1}, we collaborated with consultants highly experienced in assistive technology and multimedia accessibility for BLV users. Specifically, we engaged two blind accessibility experts: a scientist specializing in accessible technology Research \& Development (R\&D), and a professional quality control (QC) specialist for commercial AD production. Their continuous involvement directly shaped the framework's core dimensions of AD quality. The blind QC specialist further contributed by composing our corpus and developing training materials to familiarize raters with AD concepts. This collaborative process revealed a critical gap: although professional guidelines treat \textit{formatting} dimensions such as delivery method and timing as integral to practice, existing evaluation approaches rarely measure them systematically, concentrating instead only on \textit{content} dimensions \cite{2024DescribedDCMP}. These formatting aspects dictate how descriptions integrate into the audiovisual flow and are essential to the viewing experience of BLV audiences.

To answer \textbf{RQ2}, we conducted a study in which three experts first applied the framework developed to establish ground-truth ratings on a corpus of volunteer-authored and AI-generated ADs for full-length videos, without knowing which descriptions came from which source. Additional respondents then completed the same rating task: 4 human raters with accessibility-related experience (distinct from the ground-truth expert panel), 21 novice human raters, and 19 state-of-the-art VLMs. Applying IRT to these ratings enabled us to examine not only how closely VLMs approximate expert judgments relative to human raters, but also how their proficiency varies with item difficulty---nuances that aggregate metrics obscure.

As VLMs are increasingly deployed to produce and evaluate accessibility content, the field requires rigorous methods to assess whether these systems meet community standards. \textbf{Our primary contribution is a comprehensive methodological workflow for evaluating VLM and human performances on subjective quality assessment tasks in accessibility contexts.} Beyond this specific use case, the workflow is designed to be generalizable, offering the accessibility and HCI communities a practical, replicable method for assessing rater proficiency. This is a timely contribution as the field seeks robust tools to understand the growing capabilities and limitations of VLMs in accessibility work.

Our work is among the first in the accessibility and AI communities to evaluate audio description quality on full-length, uninterrupted videos (extending beyond short, isolated clips) using a combination of VLMs and human judgments from both experts and non-experts. The development and application of this workflow yields the following secondary contributions:
\begin{itemize}
    \item Empirical findings from a mixed-methods study involving human raters and state-of-the-art VLMs. By applying IRT to model rater proficiency and item difficulty on the same scale, we reveal precisely where VLMs approximate expert judgments and where they diverge.
    \item Design implications for AD rating frameworks: while our framework itself requires further psychometric validation at larger scale and with BLV users, our consultations revealed that formatting dimensions are crucial yet overlooked in prior evaluations \cite{Papineni2001BLEU, Banerjee2005METEOR:Judgments, Anderson2016SPICE:Evaluation}. We argue the community should move beyond one-dimensional constructs for evaluation in accessibility contexts.
    \item Strategies for scalable, hybrid quality control systems that combine VLM efficiency with the diagnostic depth of human feedback, supporting accessibility practitioners, volunteer-driven AD platforms, and the blind and low-vision users who depend on AD quality.
    
\end{itemize}

\section {Related Work}
\subsection{Volunteer and AI-Generated AD}

Audio description has been shown to improve comprehension, satisfaction, and engagement for BLV audiences \cite{Schmeidler2000AddingDifference, Naraine2018ImpactsStudy}. Foundational works such as Snyder's \textit{The Visual Made Verbal} \cite{Snyder2014}, along with guidelines from organizations including the Described and Captioned Media Program (DCMP) \cite{2024DescribedDCMP}, the American Council of the Blind's Audio Description Project \cite{ACBADP}, and the National Center for Accessible Media (NCAM) at WGBH Educational Foundation \cite{NationalCenterforAccessibleMedia2017AccessibleGuidelines}, establish conventions for creating high-quality AD in educational, broadcast, and film contexts. Commercial providers such as 3Play Media have further incorporated these practices into professional accessibility workflows \cite{3PlayMedia2020AudioGuidelines}, while technical standards such as WCAG 2.1 Success Criterion 1.2.5 formalize AD requirements for prerecorded video \cite{WCAG21}. Quality assurance typically involves multi-stage review: scripts are drafted and peer-reviewed, editorial checks are applied, recordings are tested for clarity and synchronization, and in some cases pilot testing with BLV users is conducted \cite{PackerAnGuidelines}. These processes make professional AD the gold standard for quality, but remain descriptive and guideline-driven, relying on expert oversight rather than quantitative assessment protocols.

To expand access beyond commercial and broadcast domains, crowdsourced platforms such as LiveDescribe and YouDescribe adapt professional guidelines for non-experts \cite{BranjeLiveDescribe:Description, YouDescribeYouDescribe.Https://www.youdescribe.org/, Jiang2022Co-DesigningWriters}. While such community-driven efforts have made AD more widely available, their outputs are rarely reviewed against accessibility standards, resulting in variability in accuracy and clarity. Prior studies of volunteer describers confirm these challenges: without guidance, novices introduce timing and placement errors that reduce usability \cite{Pavel2020Rescribe:Descriptions}, and produce lower-quality descriptions across multiple dimensions even with collaborative feedback \cite{Natalie2021TheDescriptions}. In one comparison, ADs authored by novices reached only 60\% of the overall quality of professional ADs, with notably mismatched volume \cite{Nakajima2024ProfessionalInteractions}. Beyond quality concerns, volunteer labor is strained by the explosive growth of video content on platforms like YouTube, TikTok, and Instagram, where uploads far exceed the capacity of manual description. 

To address this bottleneck, researchers have explored automation. Early automatic video description systems largely followed an encoder--decoder paradigm \cite{Aafaq2019}, where a visual model encodes frames into feature representations that a language model decodes into natural-language descriptions. More recent approaches have incorporated multimodal narration techniques \cite{Wang2021TowardVideos} and leveraged LLMs and VLMs to generate more scene-aware descriptions \cite{Chu2024LLM-AD:System, ye2024mmad, VanDaele2024MakingSummaries}. While AI systems can generate descriptions quickly at scale, persistent quality problems remain. Automated descriptions may confuse or invent characters and objects \cite{Bergin2025AutomatingAlgorithms, li2025vidhalluc}, deliver narration that is cluttered and at the wrong time, disrupting the viewing experience \cite{Wang2021TowardVideos}. Moreover, despite strong scores on captioning benchmarks, their outputs diverge from human judgments of accessibility-critical qualities such as completeness and coherence \cite{Rohrbach2017MovieDescription, Hu2024FIOVA}.

Hybrid human-AI systems that combine automated generation with human revision and on-demand question answering show promise, with BLV users reporting preferences for these approaches over AI-only alternatives \cite{Yuksel2020Human-in-the-LoopUsers, Stangl2023TheVideos}. However, quality concerns persist even within these hybrid systems. Novice describers working with AI-generated drafts rated the accuracy of the AI text as only \textit{``neutral''} \cite{Yuksel2020Human-in-the-LoopUsers}, and studies of tandem AD systems have noted ongoing algorithmic inaccuracies, including missing descriptions of salient visual content and misrepresentations of on-screen events \cite{Stangl2023TheVideos}.

The absence of quantitative assessment is especially stark for volunteer-authored and AI-generated AD, where growing volumes of content lack the rigorous expert-driven review of professional practice. Our work addresses this gap by piloting a rating framework toward the evaluation of these emerging forms of AD.

\subsection{Efforts to Evaluate AD Quality}

Evaluating the quality of audio descriptions is essential for ensuring their usability, particularly for BLV audiences who rely on AD for comprehension and engagement. Most evaluation efforts fall into two main tracks: automatic metrics and human-centered judgments.

Datasets such as VALOR \cite{Liu2025VALOR:Dataset}, YouCook2 \cite{Zhou2017TowardsVideos}, and VATEX \cite{Wang2019VATEX:Research} support AI-generated AD research, with performance typically reported using Natural Language Processing (NLP) metrics like BLEU \cite{Papineni2001BLEU}, METEOR \cite{Banerjee2005METEOR:Judgments}, and CIDEr \cite{vedantam2015cider}. Content-based approaches like SPICE \cite{Anderson2016SPICE:Evaluation} extend this by parsing scene graphs to compare objects, attributes, and relationships semantically. While these metrics are efficient, objective, and reproducible, they were developed for general captioning tasks. As a result, they privilege surface similarity and align poorly with how humans actually judge description quality \cite{Aafaq2019}. In particular, they fail to capture accessibility-critical qualities such as prioritization and temporal alignment---factors that determine whether BLV audiences experience AD as supportive or disruptive.

For both AI-generated and volunteer-authored AD, evaluation typically relies on human participants. Existing approaches often use overall quality ratings (e.g., Likert scales on understanding or satisfaction) \cite{Bodi2021AutomatedUsers}, or comparisons against a no-AD baseline, where participants flag confusion while watching without and then with generated descriptions \cite{Wang2021TowardVideos}. Such comparisons show that descriptions help, but not how good they are or where they fail. These methods also face persistent challenges. First, feedback from sighted participants, especially when gathered via crowdsourcing marketplaces like Amazon Mechanical Turk, suffers from quality issues, including inattentive workers, bots, and inconsistent responses \cite{Karpinska2021TheGeneration, Webb2024TooTurk}. Second, recruiting BLV participants remains difficult due to limited population size, reliance on advocacy organizations, accessibility logistics, and participant fatigue \cite{mack2021accessibility}. 

Beyond these methodological challenges, BLV viewers have diverse preferences for content and delivery \cite{natalie2024audio, JiangCrescentiaJungMahikaPhutaneItsScenarios}. While supporting this customization is essential, offering choices is only meaningful if the underlying descriptions are fundamentally well-executed. Therefore, before we can tailor AD to specific preferences, we first need a reliable, standardized way to assess baseline quality.

Recent work such as VideoA11y \cite{Li2025VideoA11y:Description} marks an important step forward by combining accessibility-informed evaluation with standard NLP metrics. Their framework included four custom dimensions, \textit{descriptive, objective, accurate, and clear}, alongside six standard captioning metrics, offering a more comprehensive view of AD quality than either approach alone. Their evaluation, conducted with 347 sighted participants, 40 BLV participants, and 7 professional describers, offered breadth and external validity but was resource-intensive and difficult to replicate. Their analyses were further limited to benchmark datasets such as VALOR, YouCook2, and VATEX, which contain only isolated 10–15 second clips. 

We address these limitations by extending the scope of AD evaluation in three critical directions. First, we extend evaluation to full-length AD, ranging from 1:30 to 5:05 minutes. Each video represents complete, uninterrupted content, where continuity and delivery are critical. Second, we introduce \textit{formatting} dimensions, delivery method and timing, which no prior framework has addressed despite their importance to usability. Third, we explore whether VLMs can serve as evaluators, making AD quality assessment more sustainable and scalable by applying Item Response Theory to model rater ability and item difficulty.

\subsection{LLM/VLM as Evaluators and the role of Item Response Theory}

As research on automated AD generation grows, scholars have also begun to ask whether LLMs and VLMs can serve not only as generators but also as evaluators. In NLP, the LLM-as-a-Judge paradigm has been tested in tasks such as summarization, translation, and dialogue, where model-based rankings often correlate with human preferences \cite{ZhengJudgingArena, Chiang2023CanEvaluations}. Multimodal extensions such as Prometheus-Vision demonstrate that VLMs can deliver fine-grained judgments of captions and visual explanations closely aligned with human ratings \cite{Lee2024Prometheus-Vision:Evaluation}. Within HCI, researchers frame models less as replacements and more as collaborators in evaluation workflows. EvalAssist, for example, integrates LLMs into interactive processes that reduce evaluator effort while preserving alignment with expert assessments \cite{Ashktorab2025EvalAssist:LLM-as-a-Judge}, underscoring a growing interest in hybrid strategies that combine human expertise with machine scalability. 

Despite momentum around model-as-judge frameworks, most evaluations still rely on simple validation measures, accuracy against majority vote, correlation with human scores, agreement rates, or win-rate comparisons \cite{ZhengJudgingArena, Chiang2023CanEvaluations, Lee2024Prometheus-Vision:Evaluation}. While useful, these metrics collapse complex judgments into a single number, overlooking variation in rater reliability, prompt sensitivity, and task difficulty. This is especially limiting for accessibility tasks like evaluating AD, where quality is multi-dimensional and a single aggregate cannot reveal which aspects of a description support or hinder comprehension for BLV audiences.

A more robust evaluation instead models rater ability and task difficulty jointly, rather than reducing performance to a single aggregate score. Related work in crowdsourcing has explored similar approaches by inferring worker reliability and item characteristics from noisy annotations. Whitehill et al.~\cite{Whitehill2009} modeled annotation quality as the interaction between per-annotator ability and per-item ease. Later work extended this to multidimensional annotator competence~\cite{Welinder2010} and to Bayesian joint inference of ability, difficulty, and the correct answers themselves~\cite{Bachrach2012}. Item Response Theory, an application of latent variable modeling, offers a principled framework for analyzing these interactions. Originally developed for educational testing, IRT treats \textit{items} as test questions and \textit{respondents} as students \cite{FundamentalsTheory.}. In our setting, items are audio descriptions and respondents are human and VLM raters, allowing us to estimate which descriptions are harder to evaluate and which evaluators most closely align with expert ground truth, even under partial agreement. 

We build upon these psychometric foundations to propose a comprehensive workflow for evaluating
AI raters. IRT has been adapted in HCI to handle heterogeneity in inspectors, improve questionnaire analysis, and calibrate subjective judgments \cite{Sauro2009ComparisonQuestionnaires, Schmettow2008HeterogeneityProcess}, and more recently in natural language generation evaluation to reveal systematic differences in human rater reliability \cite{Kelly2023CapturingApproach}. However, these methods have not yet been integrated into a unified workflow to assess hybrid human–AI evaluation ensembles. To the best of our knowledge, no prior work has applied this workflow in accessibility contexts, specifically the assessment of VLMs as evaluators of AD quality. By doing so, our study extends this psychometric method to a new domain, providing a richer account of both evaluator ability and AD difficulty, and advancing hybrid human–AI evaluation workflows in accessibility.

\section{Audio Description Corpus}
Stage 1 of our workflow (Fig.~\ref{fig:experimental_workflow}) assembles the corpus of audio descriptions to be rated: full-length videos paired with volunteer-authored and AI-generated ADs. These reflect the two dominant sources of AD at scale, and neither receives systematic quality assessment.

\subsection{Video Selection}
\label{sec:videos-selection}
\begin{figure*}[htbp]
  \centering

  \begin{catbox}
    \textbf{Entertainment}\par\vspace{0.3em}
    \begin{subfigure}[t]{0.23\linewidth}
      \centering
      \includegraphics[width=\linewidth]{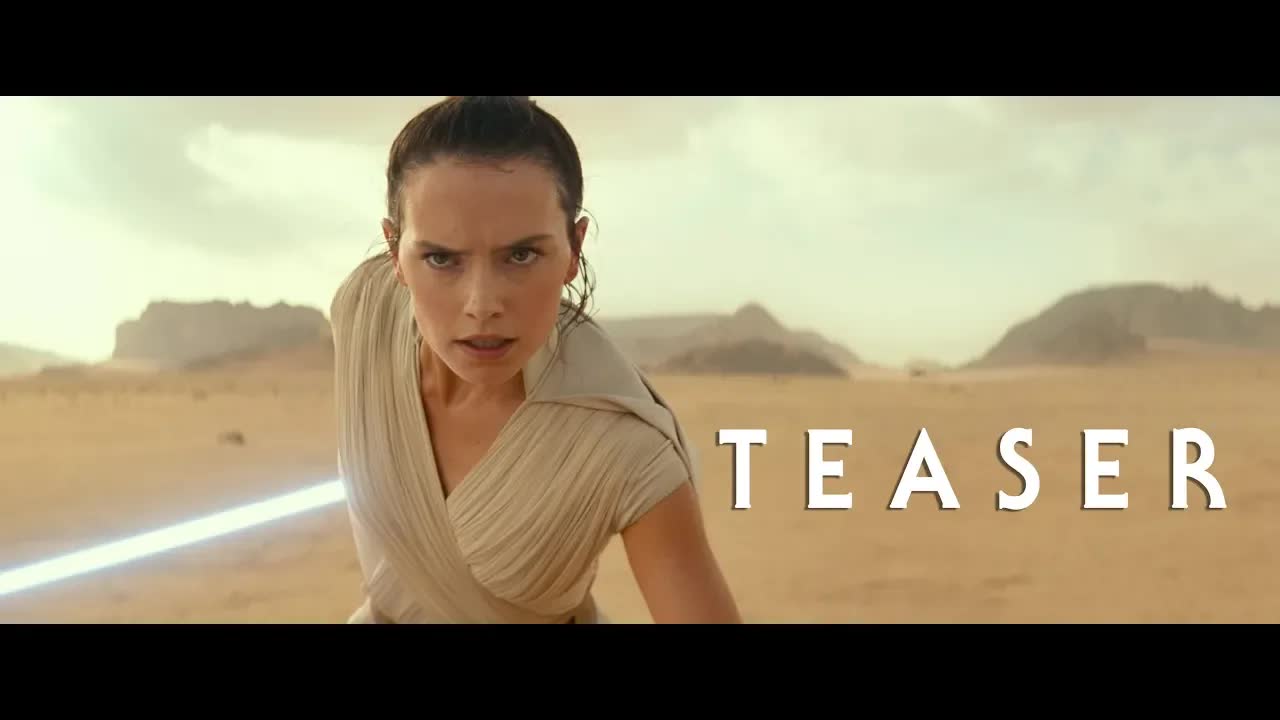}
      \caption{Star Wars: The Rise of Skywalker – Teaser(2:04)}
      \Description{Star Wars: The Rise of Skywalker – Teaser (2:04): a young woman with brown hair in a tan tunic stands in a desert landscape holding a blue lightsaber, with the word "TEASER" overlaid.}
    \end{subfigure}\hfill
    \begin{subfigure}[t]{0.23\linewidth}
      \centering
      \includegraphics[width=\linewidth]{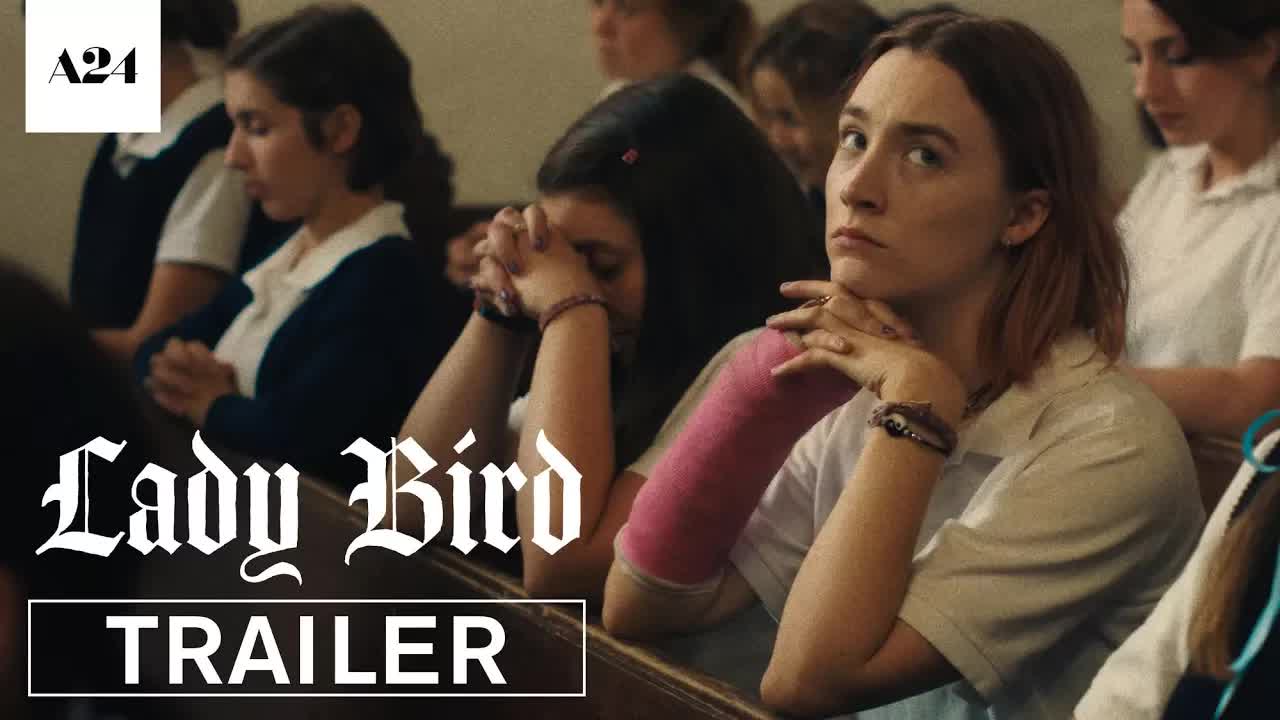}
      \caption{Lady Bird | Official Trailer HD | A24 (2:34)}
      \Description{Lady Bird | Official Trailer HD | A24 (2:34): a group of people sit together in what appears to be a church pew, with the "Lady Bird" title and "TRAILER" text displayed.}
    \end{subfigure}\hfill
    \begin{subfigure}[t]{0.23\linewidth}
      \centering
      \includegraphics[width=\linewidth]{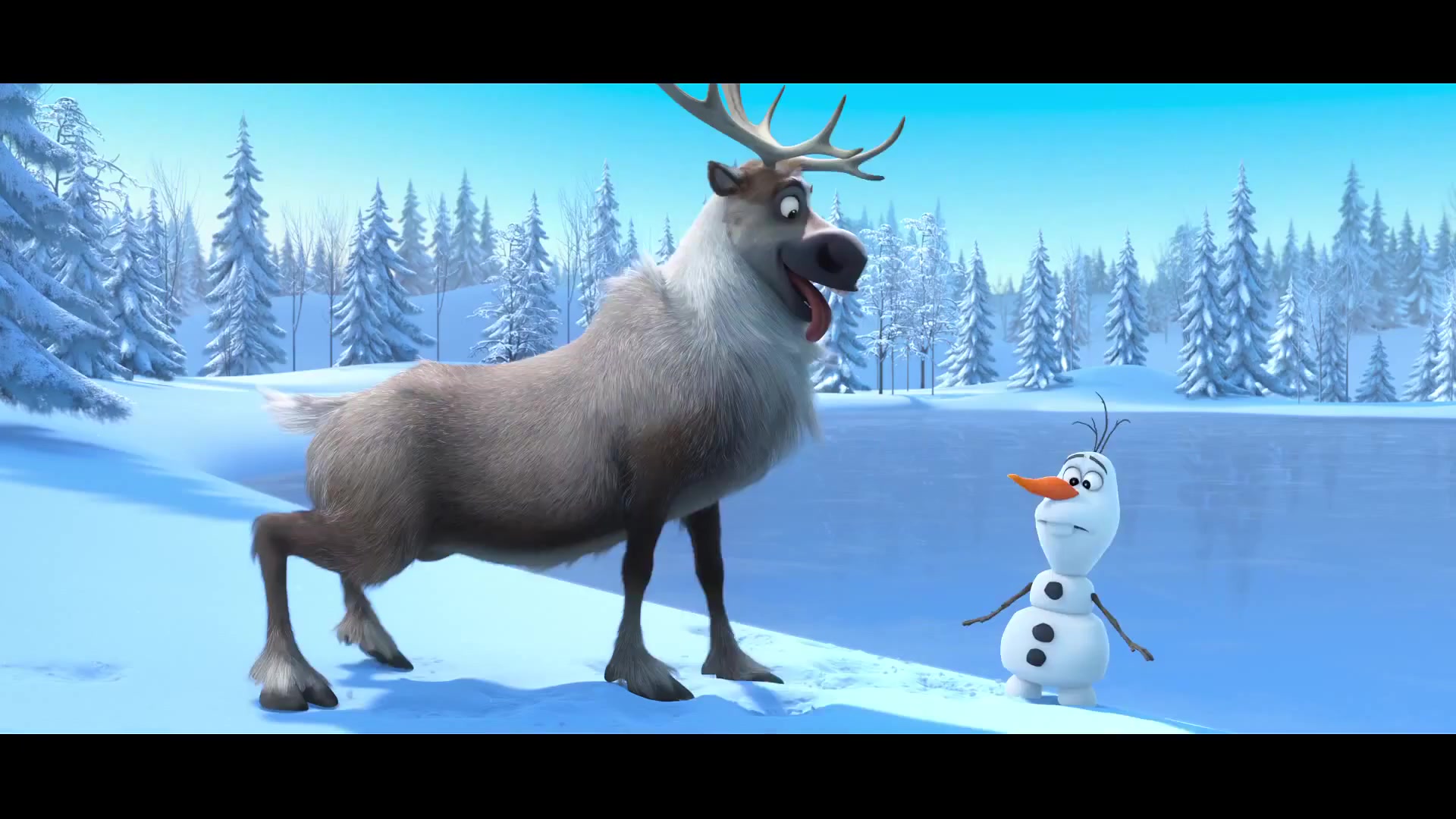}
      \caption{Frozen Teaser (2013) - Disney Animated Movie (1:30)}
      \Description{ Frozen Teaser (2013) – Disney Animated Movie (1:30): an animated reindeer stands in a snowy landscape with a small snowman nearby.}
    \end{subfigure}\hfill
    \begin{subfigure}[t]{0.23\linewidth}
      \centering
      \includegraphics[width=\linewidth]{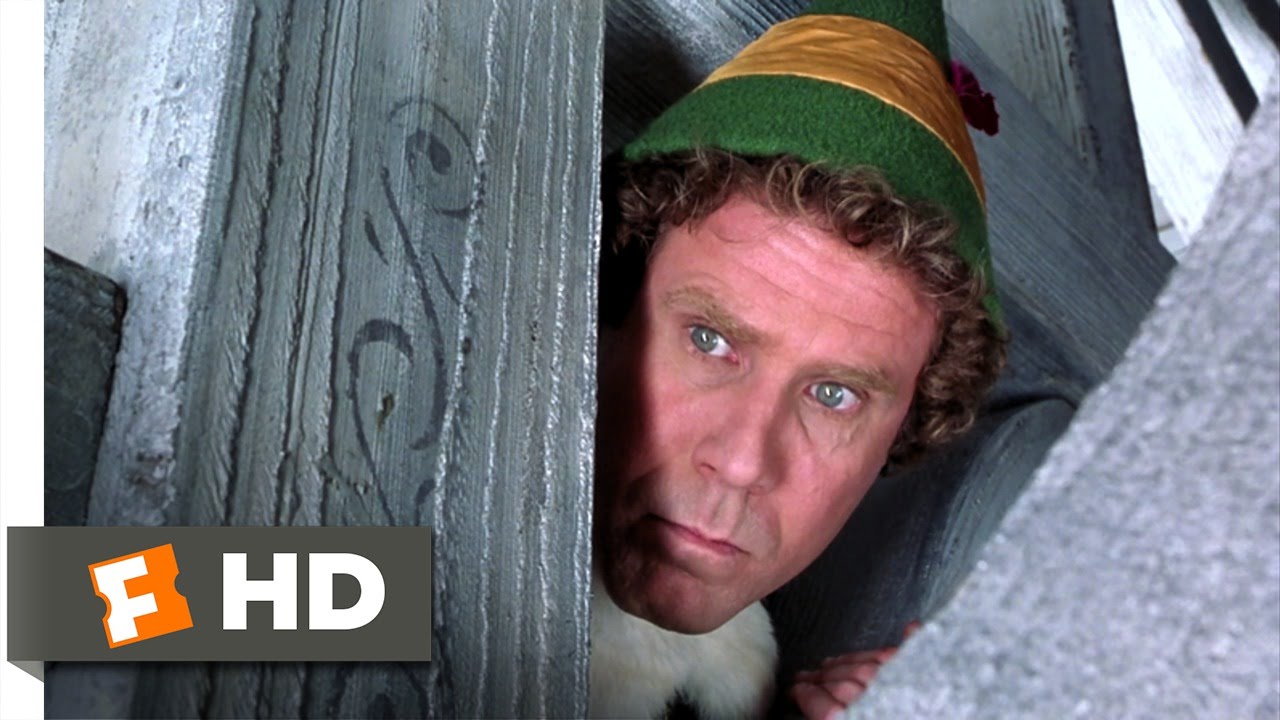}
      \caption{Elf Clip - Buddy Realizes He's Human (2003) (1:37)}
      \Description{ Elf Clip – Buddy Realizes He's Human (2003) (1:37): a close-up of a man in an elf costume with pointed ears peeking around a wooden structure, marked "HD."}
    \end{subfigure}
  \end{catbox}

  \vspace{0.8em}

   \begin{catbox}
    \textbf{How-to \& Style}\par\vspace{0.3em}
    \begin{subfigure}[t]{0.30\linewidth}
      \centering
      \includegraphics[width=\linewidth]{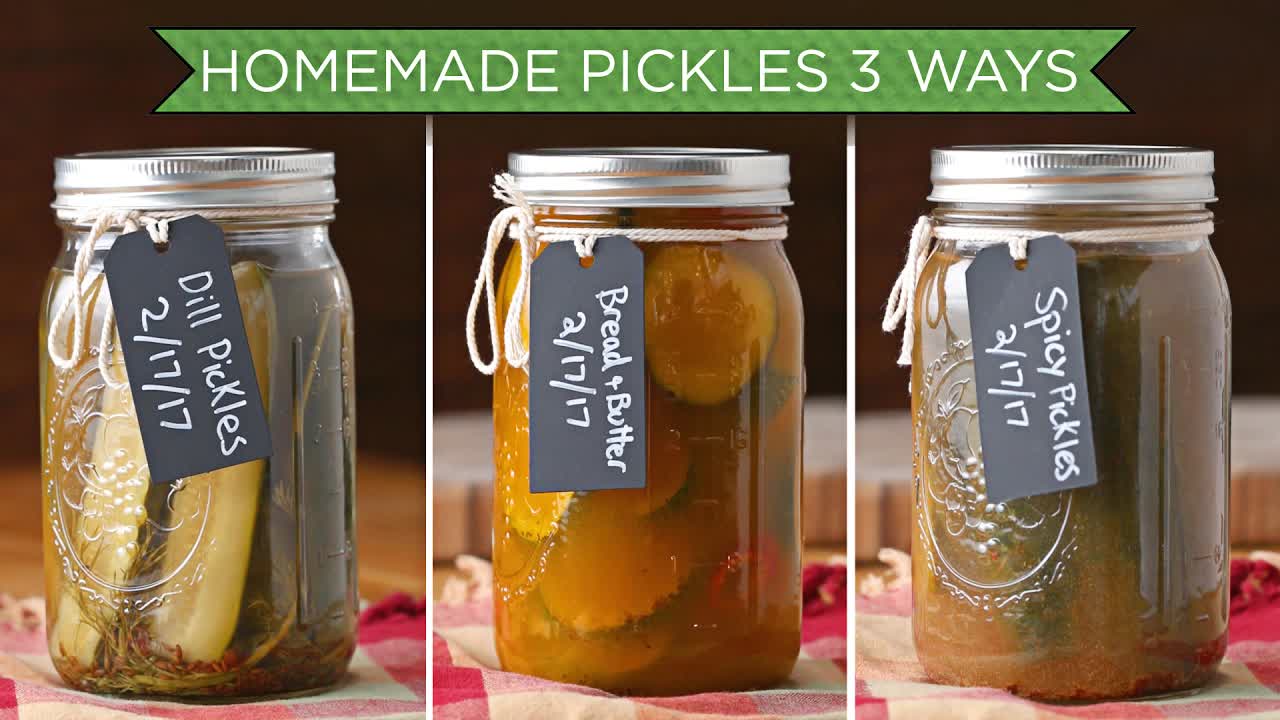}
      \caption{3 Ways to Make Homemade Pickles (2:23)}
      \Description{3 Ways to Make Homemade Pickles (2:23): three glass jars of pickles on a wooden surface with a green banner reading "HOMEMADE PICKLES 3 WAYS." }
    \end{subfigure}\hfill
    \begin{subfigure}[t]{0.30\linewidth}
      \centering
      \includegraphics[width=\linewidth]{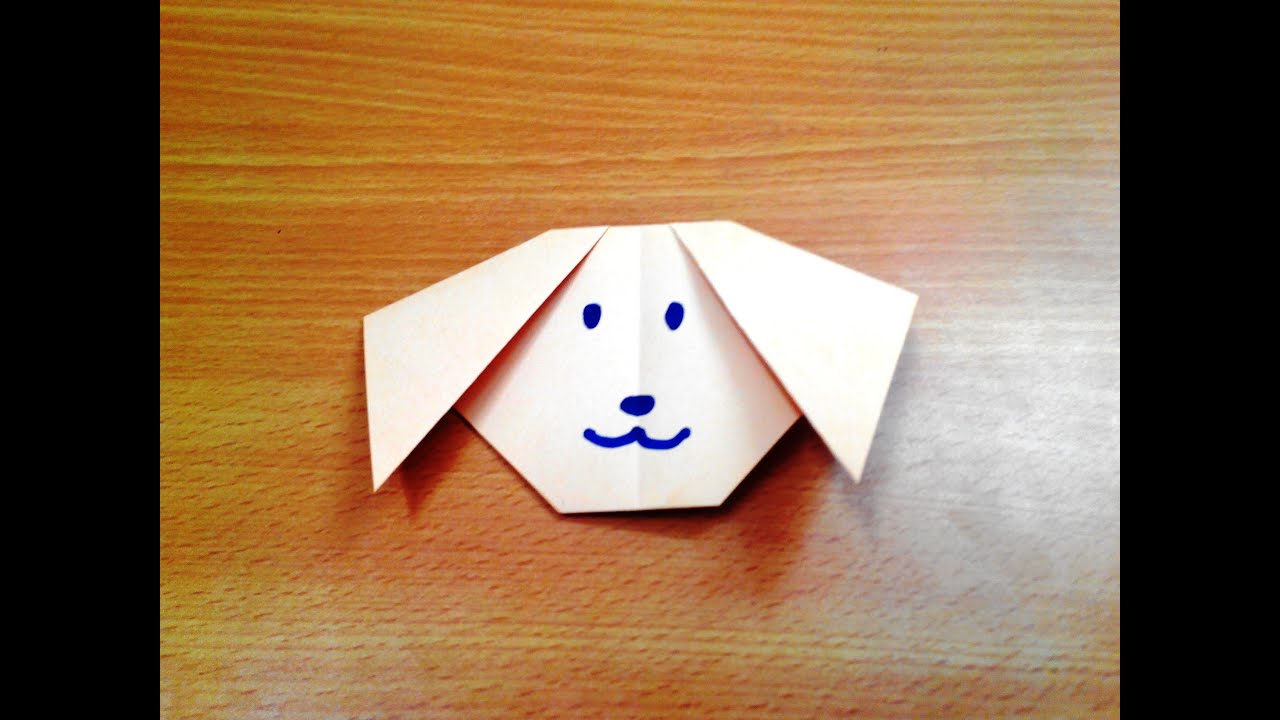}
      \caption{How to Make an Origami Dog Face (2:47)}
      \Description{How to Make an Origami Dog Face (2:47): a folded white paper origami dog face with a drawn-on nose and eyes sitting on a wooden surface.}
    \end{subfigure}\hfill
    \begin{subfigure}[t]{0.30\linewidth}
      \centering
      \includegraphics[width=\linewidth]{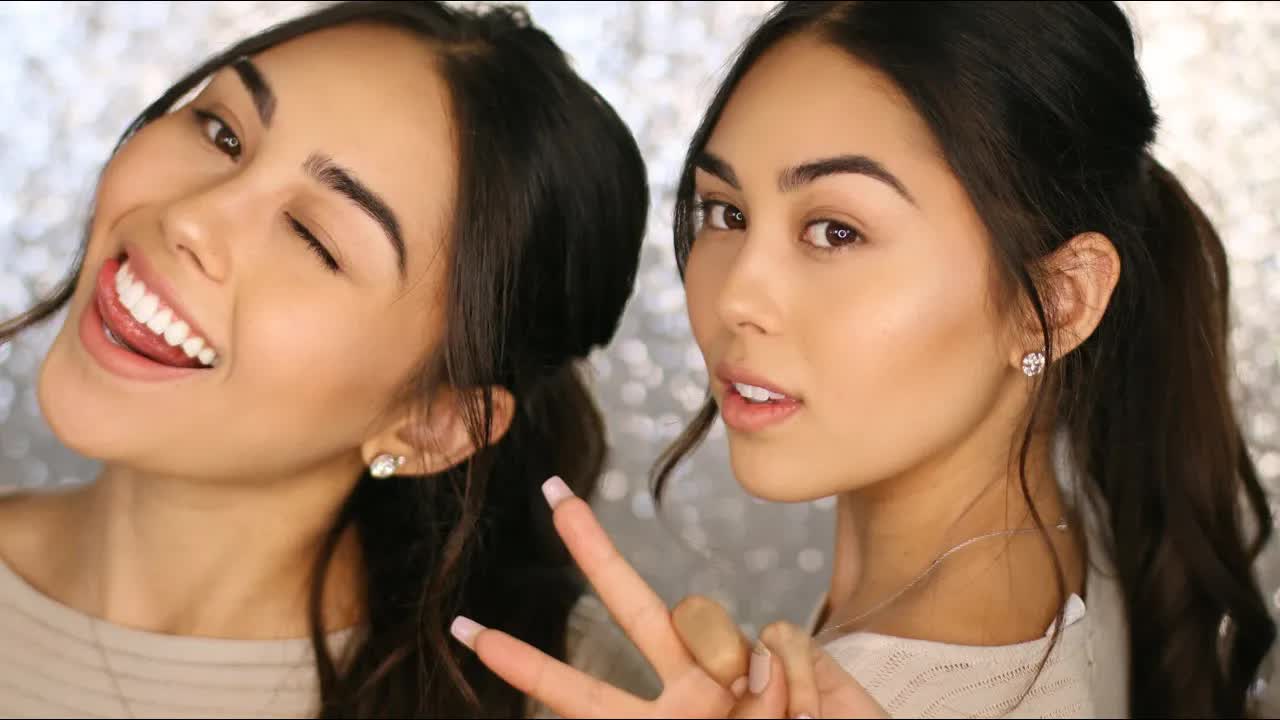}
      \caption{Quick and Easy 5-Minute Makeup Tutorial (5:01)}
      \Description{Quick and Easy 5-Minute Makeup Tutorial (5:01): two young women with dark hair posing together, one making a playful hand gesture near her face.}
    \end{subfigure}
  \end{catbox}

  \vspace{0.8em}

  \begin{catbox}
    \textbf{Education}\par\vspace{0.3em}
    \begin{subfigure}[t]{0.30\linewidth}
      \centering
      \includegraphics[width=\linewidth]{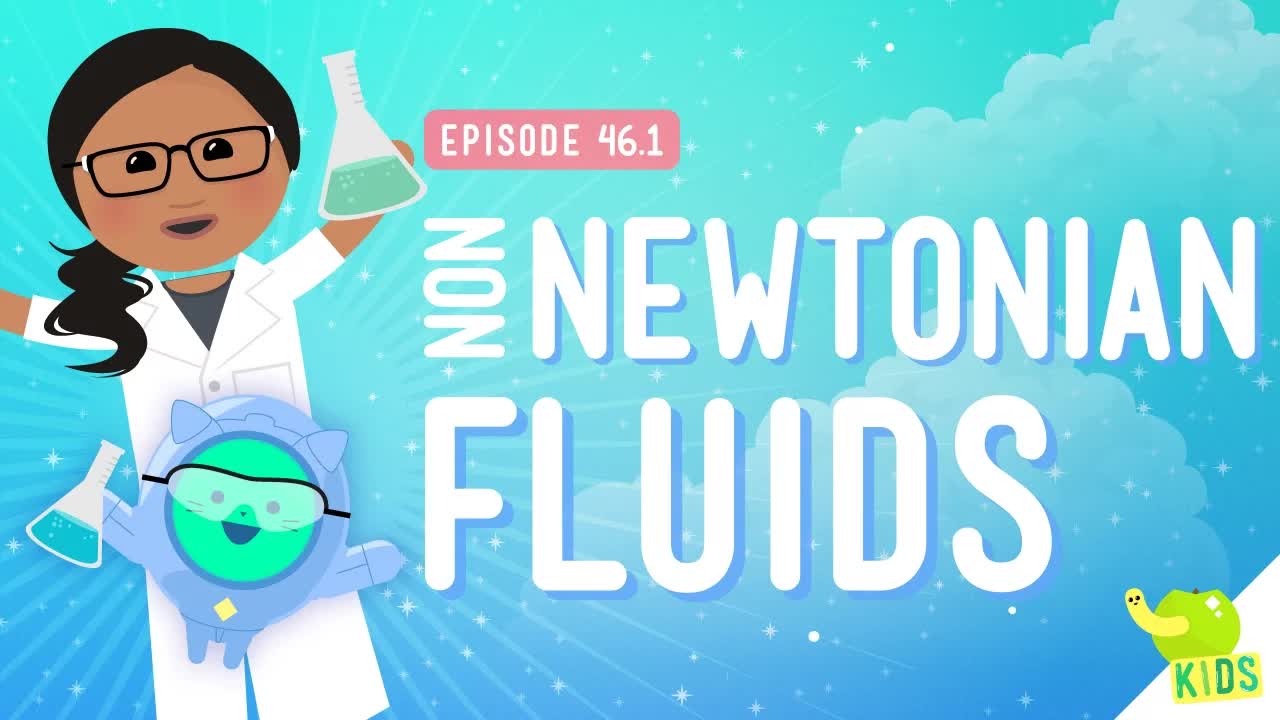}
      \caption{Non-Newtonian Fluids: Crash Course Kids (4:20)}
      \Description{Non-Newtonian Fluids: Crash Course Kids (4:20): an animated thumbnail showing a cartoon character with glasses against a blue background, with large yellow text reading "NON NEWTONIAN FLUIDS."}
    \end{subfigure}\hfill
    \begin{subfigure}[t]{0.30\linewidth}
      \centering
      \includegraphics[width=\linewidth]{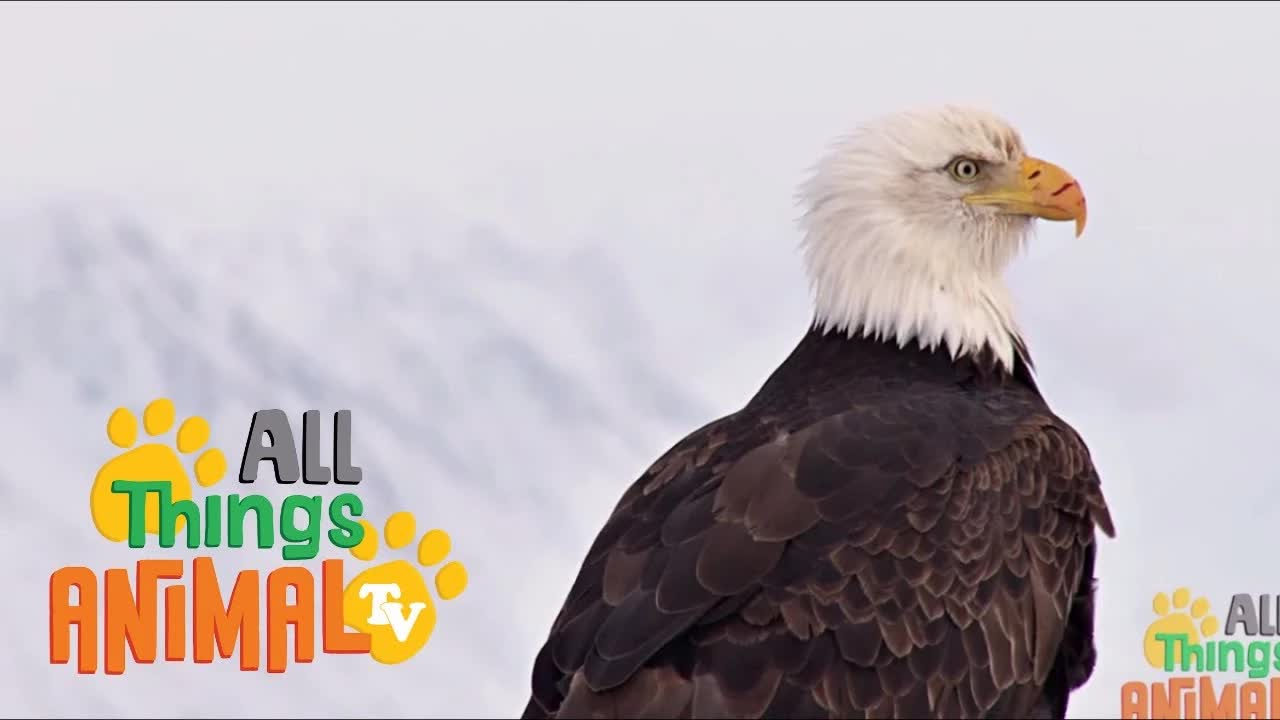}
      \caption{Bald Eagle | Animals for Kids | All Things Animal TV (3:11)}
      \Description{Bald Eagle | Animals for Kids | All Things Animal TV (3:11): a close-up of a bald eagle with a white head and yellow beak against a blue sky, with an "All Things Animal TV" logo.}
    \end{subfigure}\hfill
    \begin{subfigure}[t]{0.30\linewidth}
      \centering
      \includegraphics[width=\linewidth]{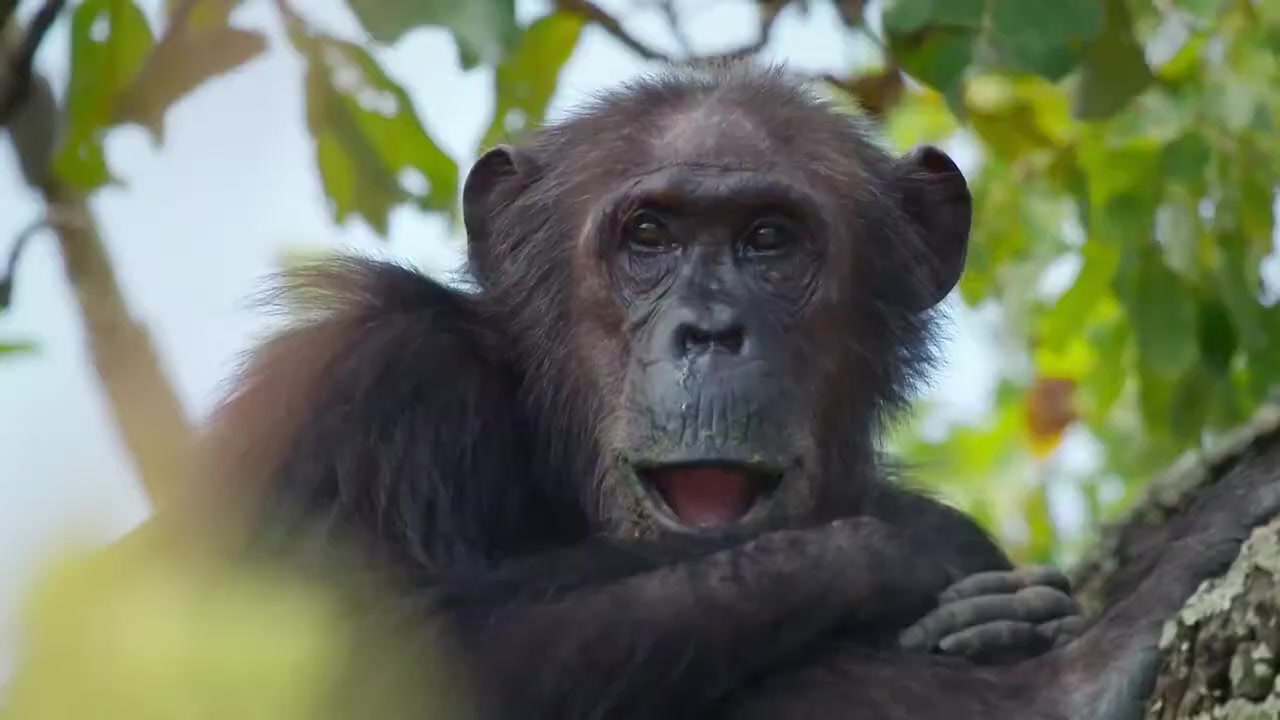}
      \caption{jane goodall (2:43)}
      \Description{jane goodall (2:43): a chimpanzee sitting in green foliage, looking toward the camera.}
    \end{subfigure}
  \end{catbox}

  \vspace{0.3em}
  \caption{Ten YouTube videos used in the evaluation, grouped by category: Entertainment (top), How-to \& Style (middle), and Education (bottom). Subcaptions show the YouTube titles with its video length (all thumbnails used under Fair Use).} 
\Description{The figure shows the thumbnails of 10 YouTube videos used under Fair Use for evaluation, with their titles and video lengths on the captions. They are grouped by categories of Entertainment, How-to \& Style, and Education. Starting with the Entertainment category which is located at the top, there are 4 videos from the left: (a) Star Wars: The Rise of Skywalker – Teaser, (b) Lady Bird Official Trailer, (c) Frozen Teaser, (d) Elf Clip. How-to \& Style category is in the middle with the 3 videos: (e) 3 Ways to Make Homemade Pickles, (f) How to Make an Origami Dog Face, (g) Quick and Easy 5-Minute Makeup Tutorial. Lastly, Education category is located at the bottom with the 3 videos: (h) Non-Newtonian Fluids: Crash Course Kids, (i) Bald Eagle Animals for Kids All Things Animal TV, (j) jane goodall.}

  \label{fig:study-videos}
\end{figure*}

To create a corpus that was both varied in quality and genre, we adopted a three-step selection process.

First, we drew from a broad pool of videos containing volunteer-authored audio descriptions publicly available on YouDescribe \cite{YouDescribeYouDescribe.Https://www.youdescribe.org/}. Each video is rated by the community on a 1–5 star scale (5 being best). We selected videos ranging from 2 to 5 stars to capture variability in quality, using these ratings only as an initial filter since the raters are anonymous and their criteria cannot be verified.

Second, an accessibility consultant with experience working with BLV users and training new audio describers conducted a pre-screen of this pool. The consultant sorted videos into stronger or weaker AD based on overall impression. This sorting was not used in later analyses; instead, this step helped ensure our final sample captured a range of descriptive demands and perceived quality.

Third, drawing on the consultant’s pre-screening, we further curated ten videos spanning multiple genres and audiences. Entertainment trailers (\textit{Star Wars}, \textit{Lady Bird}) required maintaining narrative continuity amid dense visuals and tonal shifts. Family-oriented films (\textit{Elf}, \textit{Frozen}) demanded age-appropriate language while conveying humor and action. How-to \& Style clips (\textit{3 Ways to Make Homemade Pickles}, \textit{Quick and Easy 5-minute Makeup tutorial}, \textit{How to Make an Origami Dog Face}) required stepwise clarity to describe fine-grained hand movements. Education videos ranged from the rapid-fire style of \textit{Crash Course for Kids} to the child-oriented \textit{Bald Eagle | Animals for Kids} and the slower-paced \textit{Jane Goodall} at \textit{Gombe}, which required integrating factual narration with scenic content.

Following selection, a blind professional AD quality controller reviewed descriptions from a subset of these videos. She independently confirmed that the genres posed meaningfully distinct descriptive demands, and additionally surfaced consistent identification in dialogue-dense educational videos as a challenge.

Together, the crowdsourced quality signals, expert pre-screening, and genre-based curation yielded a corpus that reflects what we find to be a representative sample of real-world audio description demands from blind and low-vision audiences. It also provides a testbed for examining how human and VLM raters assess AD quality across genres, audiences, and levels of descriptive difficulty.

\subsection{Audio Descriptions}
For each of the ten videos, we prepared three audio description versions---one volunteer-authored and two VLM-generated---resulting in 30 audio descriptions.

\subsubsection{Volunteer-authored Audio Description}

Volunteer-authored descriptions were retrieved from YouDescribe, where they are publicly available alongside their source videos. The platform provides guidance on AD best practices and encourages contributors to follow these guidelines before publishing. Contributors are anonymous, so individual experience levels cannot be confirmed; the corpus instead spans a deliberate range of AD quality, established through community ratings and pre-screening by our professional describer consultant (\S\ref{sec:videos-selection}).
The recordings were transcribed with Whisper \texttt{large-v3} ($\approx$98\% accuracy) \cite{Radford2022RobustSupervision}. Our team reviewed each transcript and manually corrected transcription errors, preserving the volunteers' original wording. To anonymize speakers and standardize delivery, transcripts were resynthesized with a single synthetic voice using Google Cloud Text-to-Speech (TTS). Speech rates were adjusted to approximate the pacing of the original volunteer (e.g., faster for action-oriented trailers and slower for nature content) thereby preserving the intent of the narration while removing potential bias from vocal delivery.

\subsubsection{VLM-generated Audio Description}

\paragraph{VLM Selection}
We initially considered three state-of-the-art VLMs: Qwen2.5-VL \cite{Bai2025Qwen2.5-VLReport}, Gemini 1.5 Pro \cite{GoogleDeepMind2024GeminiReport}, and GPT-4o \cite{OpenAI2024GPT-4oCard}. At the time of development, all ranked among the top models on the Video-MME benchmark \cite{Fu2024Video-MME:Analysis}, representing both proprietary (Gemini, GPT) and open-source (Qwen) systems.

Our final corpus was generated with Qwen2.5-VL-72B-Instruct and GPT-4o, as both are widely used in recent AI-generated AD research \cite{Chu2024LLM-AD:System, Li2025VideoA11y:Description, baghel2025towards}. We excluded Gemini based on our blind consultant's evaluation. She found its outputs comparable to GPT-4o's on our videos but preferred GPT-4o despite Gemini's higher Video-MME ranking. This observation aligns with prior findings that benchmark performance on general multimodal tasks does not necessarily predict alignment with human judgments on accessibility-critical qualities \cite{Rohrbach2017MovieDescription}. 

\begin{figure*}[t]
  \includegraphics[width=\textwidth]{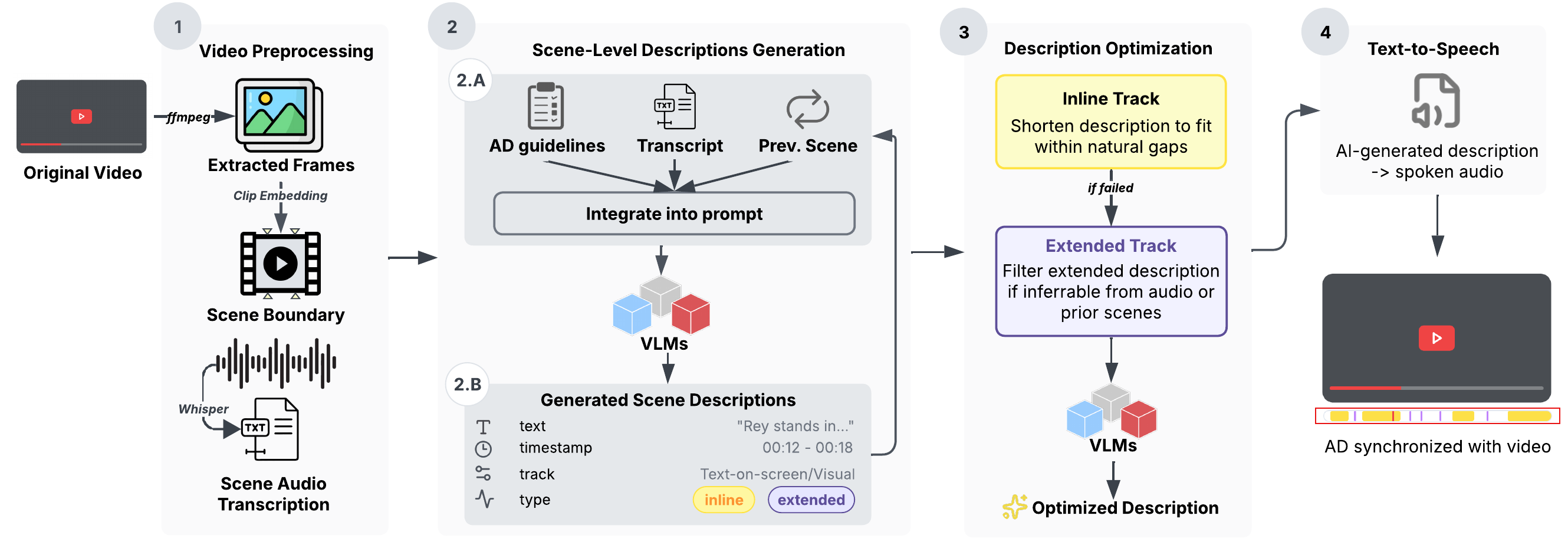}
  \caption{VLM-generated description pipeline. Videos are segmented into scenes, a VLM generates scene-level descriptions, a second VLM pass optimizes them, and the outputs are resynthesized and synchronized with the video.}
  \Description{A four-stage pipeline diagram flowing left to right, titled by stage. An original video enters stage 1, Video Preprocessing, via ffmpeg. This stage produces extracted frames, which are converted through CLIP embedding into scene boundaries, and, on a parallel path, scene audio transcription obtained by running Whisper over the audio track. Stage 2, Scene-Level Descriptions Generation, has two sub-panels. In 2.A, three inputs, AD guidelines, transcript, and previous scene descriptions, are integrated into a single prompt that is passed to the VLMs. In 2.B, the VLMs emit generated scene descriptions as structured records with four fields: text, for example ``Rey stands in\dots''; timestamp, for example 00:12 to 00:18; track, either text-on-screen or visual; and type, either inline or extended. An arrow loops from 2.B back to the previous-scene input in 2.A, so each scene is conditioned on the ones before it. Stage 3, Description Optimization, first attempts an inline track that shortens the description to fit within natural dialogue gaps; if that fails, an extended track filters out descriptions inferable from audio or prior scenes. Both pass through the VLMs to yield an optimized description. Stage 4, Text-to-Speech, converts the descriptions to spoken audio and synchronizes them with the video, shown as a video timeline marked with inline and extended segments.}
  \label{fig:vlm_pipeline}
\end{figure*}

\paragraph{Generation Pipeline}
The generation pipeline consists of four stages: scene segmentation, scene-level generation, description optimization, and text-to-speech conversion (Fig.~\ref{fig:vlm_pipeline}, 1--4).

\begin{enumerate}
\item \textit{Video Preprocessing:}
This stage involved frame extraction, scene boundary detection, and audio transcription. We sampled every 15th frame using \texttt{ffmpeg} \cite{TomarSuramya2006ConvertingFFmpeg} and computed per-frame embeddings with CLIP ViT-B/32 \cite{Radford2021CLIP}. Cosine similarity between consecutive frames identified scene boundaries (threshold 0.85), grouping visually consistent segments. Word-level timestamps were extracted from the original audio using the \texttt{whisper-timestamped} package, which applies dynamic time warping over Whisper \texttt{large-v3} cross-attention weights \cite{lintoai2023whispertimestamped, Radford2022RobustSupervision, Giorgino2009ComputingPackage}.

\item \textit{Scene-level Generation:}
Each scene was prompted with three inputs (Fig.~\ref{fig:vlm_pipeline}, 2.A): professional AD guidelines emphasizing accurate, concise narration \cite{2024DescribedDCMP, NationalCenterforAccessibleMedia2017AccessibleGuidelines, YouDescribeYouDescribe.Https://www.youdescribe.org/}, the scene transcript, and descriptions generated from previous scenes. The generated outputs are structured, containing timestamps, event type (\textit{Text-on-Screen} or \textit{Visual}), and narration mode (\textit{Inline} or \textit{Extended}) (Fig. \ref{fig:vlm_pipeline}, 2.B). Inline narration is placed within natural dialogue pauses, whereas extended narration briefly pauses playback to provide additional detail \cite{YouDescribeYouDescribe.Https://www.youdescribe.org/}.

\item \textit{Description Optimization:}
Based on consultant feedback, the pipeline prioritized inline narration to minimize playback interruptions. Extended descriptions were iteratively shortened to fit available dialogue gaps while preserving critical details. When inline delivery remained infeasible, a filtering prompt retained only extended descriptions containing necessary information not inferable from audio or prior scenes.

\item \textit{Text-to-speech:}
Optimized descriptions were synthesized using the same TTS system as the volunteer-authored descriptions, supporting both inline and extended narration.
\end{enumerate}

Qwen2.5-VL-72B-Instruct was run locally with 4-bit NF4 quantization and FlashAttention-2, while GPT-4o was accessed via the API (\texttt{gpt-4o}, August 2025). All other pipeline components were held constant. Full prompts are provided in Appendix~\ref{gen_prompts}, and the complete implementation is available on GitHub for reproducibility.\footnote{\url{https://github.com/lanad22/VideoDescribe.git}}

\section{Rating Framework}
\label{sec:framework}
Stage 2 of our workflow (Fig.~\ref{fig:experimental_workflow}) operationalizes AD quality into the measurable dimensions the workflow requires as input. We developed this rating framework in consultation with accessibility experts and blind consultants, grounding it in established professional guidelines while extending them to address gaps we identified for full-length video.

\subsection{Accessibility Consultants Insights and Motivation}

The rating framework was shaped through an iterative refinement process that grounded professional standards in real-world application. We began with the DCMP Description Quality Key \cite{2024DescribedDCMP}, the industry gold standard, before stress-testing our criteria through multi-disciplinary consultations with:
\begin{enumerate}
    \item A Professional Audio Describer with experience training and evaluating new describers' work. She emphasized that effective AD balances accurate visual depiction with strategic editorial choices, specifically regarding inline versus extended description and precise track placement.
    \item A Blind Scientist with decades of experience leading accessible technology R\&D. Drawing on his work with scientific and dialogue-heavy media, he emphasized that extended descriptions must be used strategically as they are valuable when applied selectively, but distracting if overused, making strategic implementation a key indicator of AD quality.
    \item An Assistive Technology Specialist who provides training for Teachers of Students with Visual Impairments (TVIs), with expertise in universal design and digital multimedia accessibility. She observed that the emerging criteria naturally clustered into two overarching categories: \textit{content} and \textit{formatting}. This framing crystallized a key insight: description quality depends equally on what is described and how and when descriptions are delivered.
    \item A Blind Quality Control (QC) Specialist. Reviewing sample ADs from our corpus, she frequently identified cases where accurate content was rendered unusable by delivery issues, such as narration stepping on essential dialogue, music drowning out descriptions, or awkward playback pauses. Her feedback further grounded these formatting concerns in practice.
    
\end{enumerate}

While prior research has largely focused on content quality, whether descriptions are accurate or descriptive enough, our consultants stressed that formatting is crucial to whether description is functional: poor timing disrupts comprehension \cite{Naraine2018ImpactsStudy}, and the absence of extended narration in dense contexts omits critical information \cite{3PlayMedia2020AudioGuidelines}. Despite its importance, formatting has been largely overlooked in AD evaluation research. Benchmarks such as YouCook2 \cite{Zhou2017TowardsVideos}, VATEX \cite{Wang2019VATEX:Research}, and VALOR \cite{Liu2025VALOR:Dataset} contain only short clips where delivery issues rarely surface, and most AI-based pipelines default to generating descriptions for silent sections, implicitly producing only inline-style narration \cite{ye2024mmad, Chu2024LLM-AD:System}. This narrow focus sidelines delivery concerns, even as volunteers on platforms like YouDescribe author full-length AD where extended narration and precise timing are essential \cite{YouDescribeYouDescribe.Https://www.youdescribe.org/}. 

As one consultant noted: \textit{``Audio description without careful formatting is like a paper with strong ideas but no structure.''} This insight motivated our dual-pillar evaluation model, which treats content and formatting as equally critical dimensions of AD quality.

\subsection{Rating Framework Structure and Dimensions}
The framework is organized into two categories, content and formatting, detailed below.
\subsubsection* {Content Dimensions}
The DCMP Description Quality Key identifies essential dimensions that AD must satisfy to be considered high quality \cite{2024DescribedDCMP}. We draw on four of the five keys, listed below, excluding the \textbf{Appropriate} key, which concerns avoiding offensive or inappropriate content. This exclusion reflects the structure of our evaluation rather than the importance of the criterion. Unlike the other keys, appropriateness is binary: a description either meets the standard or it does not, making it unsuitable for the graded scale the IRT model requires. Consistent with this, all descriptions in our corpus met this criterion. In future deployments, this key could serve as an initial screening filter before descriptions enter the rating workflow.

\begin{enumerate}
    \item \textbf{Accurate}: Descriptions are factually correct and error-free. 
    \item \textbf{Prioritized}: Information critical for comprehension and enjoyment should be emphasized, while less important details are minimized.  
    \item \textbf{Consistent}: Terminology, tone, and pacing should align with the program’s style and remain uniform throughout the narration.  
    \item \textbf{Equal}: Descriptions should preserve the program’s meaning and intent for equitable access without distortion or bias.  
\end{enumerate}

\subsubsection* {Formatting Dimensions}

Delivery choices, both method and track placement, are as critical as content in determining AD quality. With crowdsourced platforms now supporting both inline and extended narration, describers increasingly face decisions about how to deliver descriptions effectively. To account for this, we incorporated two delivery-focused dimensions into our evaluation model, drawing directly on NCAM broadcast guidelines \cite{NationalCenterforAccessibleMedia2017AccessibleGuidelines}, DCMP standards \cite{2024DescribedDCMP}, and professional industry practices such as 3Play Media \cite{3PlayMedia2020AudioGuidelines}.

\begin{enumerate}[resume, topsep=2pt, partopsep=0pt]

\item \textbf{Strategic Use of Description Method (Inline vs. Extended)}: Professional guidelines recommend inline narration as the preferred method when natural pauses allow visual details to be conveyed without disrupting the program. Extended narration is advised when natural pauses are insufficient, for instance, in dialogue-heavy, text-heavy, noisy, or fast-cut videos where critical information would otherwise be lost.

\item \textbf{Timing and Placement}: Guidelines emphasize inserting narration as close as possible to the relevant visual action while avoiding overlap with dialogue or essential sounds. Both NCAM and DCMP recommend using natural pauses, aligning narration with the visual timeline, and where appropriate, allowing pre-description if it clarifies the scene.

\end{enumerate}

We structured these dimensions into 1–5 rating-level descriptors drawn from our blind QC specialist's professional evaluation criteria. The resulting framework is grounded in established standards and shaped by input from both sighted accessibility experts and blind professionals. It extends prior approaches by incorporating two formatting dimensions, delivery method and timing, which have been overlooked in short-clip evaluations. This combination enables assessment of full-length AD and provides the necessary foundation for the subsequent workflow stages.

\section{Rating Protocol}
Stages 3 and 4 of our workflow (Fig.~\ref{fig:experimental_workflow}) evaluate how closely human and VLM raters align with expert ground truth. Three experts first established ground-truth ratings for the corpus; human respondents and VLMs then completed the same rating task under matched conditions.

\subsection{Establishing Ground Truth Ratings}
We recruited three accessibility consultants, independent from those who contributed to the framework's development, to serve as the ground-truth expert panel. All three have extensive experience in accessibility research and community engagement, including service on accessibility advisory boards, creating and reviewing accessible media resources, and training BLV users on assistive technologies. Because factual accuracy was one of the evaluation dimensions, sighted experts were needed to verify whether descriptions faithfully represented on-screen content. While BLV users did not directly participate in this stage, accessibility consultants anchored the evaluation in BLV-centered principles and standards. Using the assessment framework, the experts rated descriptions on a 1–5 ordinal scale across all six dimensions, supported by comprehensive documentation including definitions, explanatory notes, illustrative examples, and evaluation criteria for each dimension (full materials are available in the Appendix~\ref{Appendix:framework}).

\subsubsection{Interface}

\begin{figure}[htbp]
\centering 
\begin{subfigure}[t]{0.65\linewidth} 
\centering 
\includegraphics[width=\linewidth]{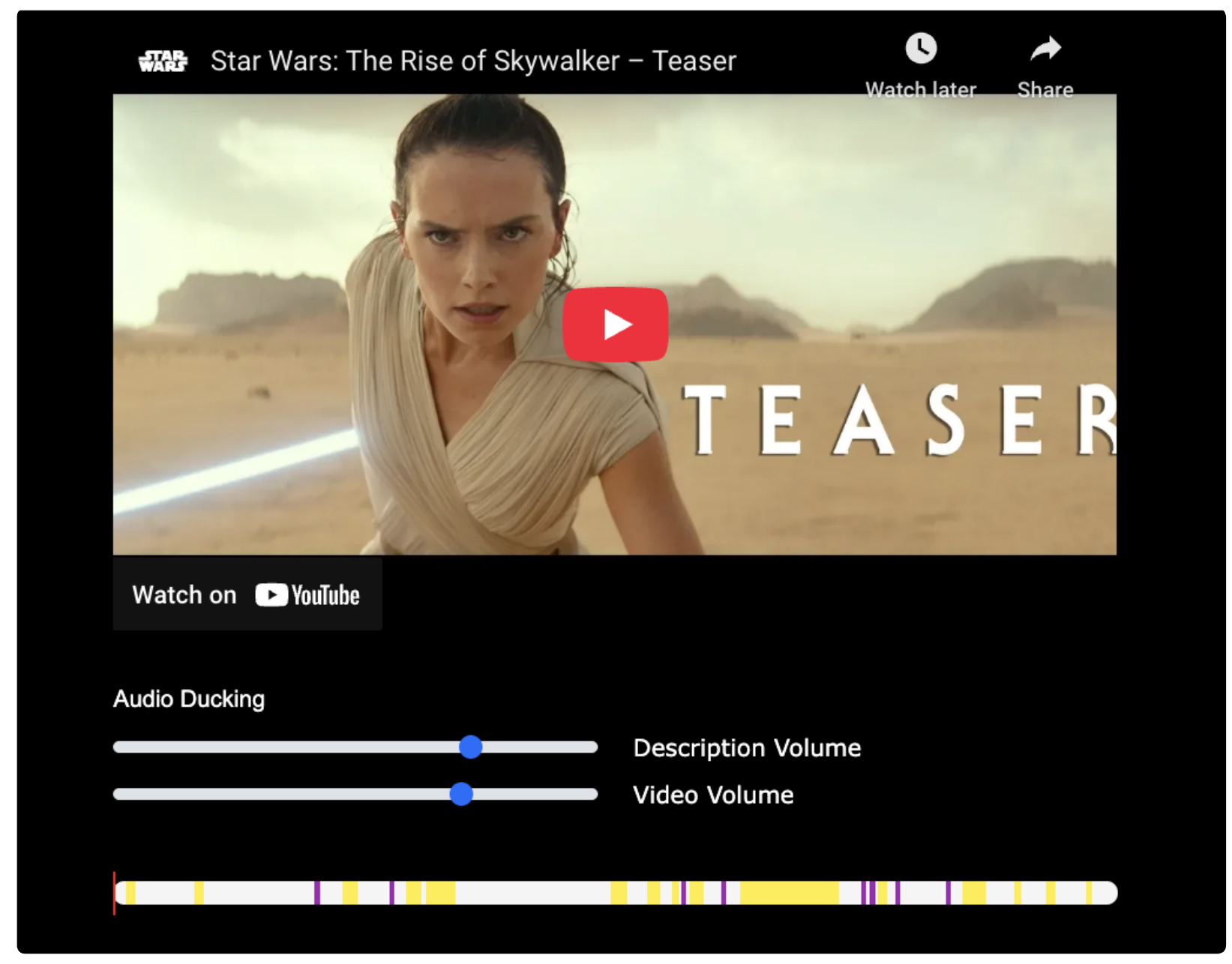} 
\Description{YouTube video player showing a Star Wars: The Rise of Skywalker teaser, with a woman holding a lightsaber in a desert setting. Below the video are two horizontal sliders for Audio Ducking: Description Volume set near the middle and Video Volume set slightly lower. Beneath the sliders is a timeline bar marked with yellow segments for inline descriptions and purple segments for extended descriptions, distributed across the video duration.} 
\caption{Interface showing inline (yellow) and extended (purple) narration markers on the timeline, with audio ducking controls.} 
\label{view-1} 
\end{subfigure} 
\hspace{0.04\linewidth} 
\begin{subfigure}[t]{0.25\linewidth} 
\centering 
\includegraphics[width=\linewidth]{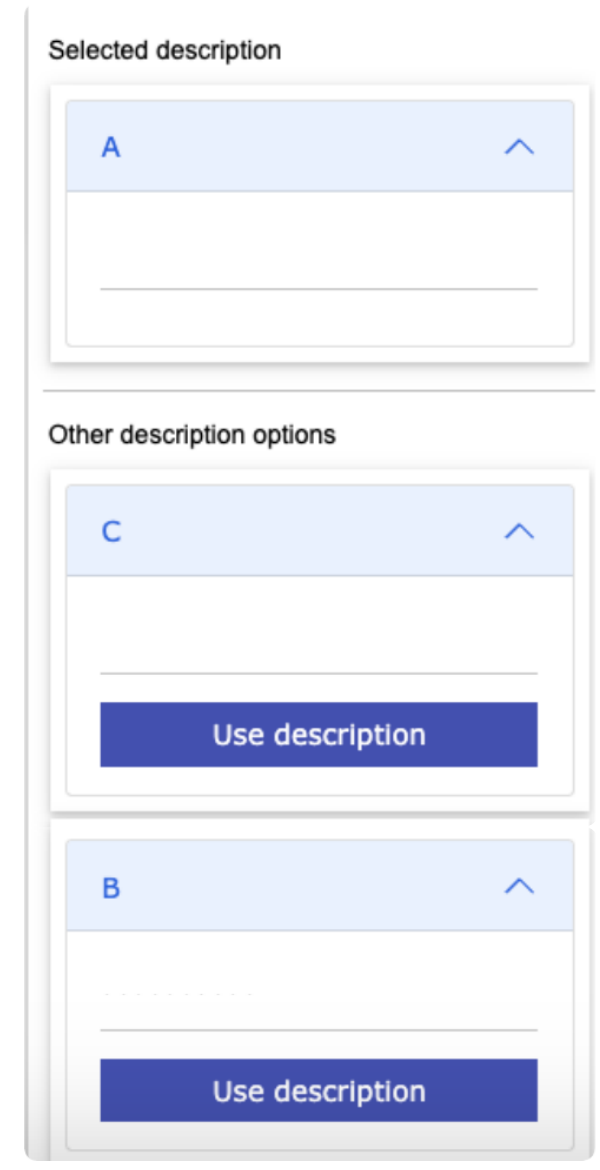} 
\Description{Vertical list of three AD version options labeled A, B, and C. Version A is shown at the top under the heading "Selected description." Versions C and B appear below under "Other description options," each with a "Use description" button for switching the active version.} 
\caption{Available AD versions.} 
\label{view-2} 
\end{subfigure} 
\caption{Viewing interface for synchronized playback of audio description with video (third-party video components used under Fair Use).} 
\Description{Two-panel figure showing the viewing interface. Panel (a) shows a YouTube video player with an embedded third-party video thumbnail, Audio Ducking sliders for Description Volume and Video Volume, and a timeline bar marked with yellow inline-narration segments and purple extended-narration segments. Panel (b) shows a vertical list of three AD version options labeled A, B, and C, with version A selected at the top and versions C and B available below, each with a "Use description" button for switching the active version.} 
\label{fig:interface} 
\end{figure}

The experts interacted with a custom-built interface designed to present audio descriptions in context with the video (Fig.~\ref{fig:interface}). The interface embedded the YouTube video and played the AD clip alongside it, the same setup as YouDescribe. Inline narration segments appeared as yellow overlays on the playback timeline, while extended narrations were shown as purple markers that briefly paused playback for longer insertions. Since YouTube source audio varies in loudness, raters could adjust the relative volume of the video and the AD, and were instructed to set the two sources to comparable levels. These controls allowed raters to clearly perceive both the content of the description and its delivery.

\subsubsection{Evaluation Procedure}
\label{sec:procedure}

For every video, three AD versions were available---one volunteer-authored and two VLM-generated---displayed at the bottom right of the video player (Fig.~\ref{view-2}). To establish a \textit{ground truth}, accessibility experts independently evaluated these descriptions under randomization at three levels: descriptions were relabeled with neutral identifiers (A–C) remapped per video, so ``A'' might denote the human-authored version in one video and a Qwen-generated one in the next; video order was randomized per expert; and within each video, version order was randomized so no system consistently appeared first or last. These measures obscured description provenance and encouraged independent evaluation.

Each expert was provided with a personalized instruction sheet specifying their randomized sequence of videos and AD versions. For each version, participants followed a link that opened the interface with the current AD displayed under \textit{``Selected description''} as seen in Fig.~\ref{view-2}. Experts then listened to the AD and completed an evaluation form rating it across six dimensions, with an optional text field for qualitative comments. 

To establish the ground truth for analysis, we applied a majority/consensus rule to these expert ratings. When two experts independently assigned the same score, that value was taken as the ground truth. In the less frequent cases where all three experts diverged, we resolved disagreement by selecting the median rating, ensuring that the reference score reflected central tendency rather than privileging any single rater. This process produced the expert reference ratings that serve as the ground truth in our study.

\subsection{Human Respondents}

We recruited 25 human respondents across two groups. The first group comprised four participants with accessibility-related experience, including work as trained audio describers, teachers of students with visual impairments, and assistive technology developers. They were selected for their ability to engage meaningfully with a six-dimension rating task across 30 audio descriptions. Because recruiting participants with accessibility expertise was costly and limited in availability (the same constraints that shaped our three-person ground-truth panel), we included four such respondents alongside a larger novice group.

The second group consisted of 21 graduate students (ages 25–35) with no self-reported experience in audio description or accessibility-related work. We developed an in-person orientation module that explained each framework dimension and walked through example ratings before the task began. To ensure clarity, the module incorporated accessible, plain-language framing contributed directly by our blind QC specialist, drawing on her experience training others in AD evaluation. An accessibility expert was present throughout the session and answered questions as raters worked.

Only the novice group completed the orientation to gain necessary context, and all raters ultimately followed the same evaluation process as the expert panel described in Section 5.1. They assessed three AD versions for each of 10 videos using our six-dimension framework. The same anonymization and randomization procedures were applied to prevent ordering effects and ensure that respondents could not identify the source of each AD. Ultimately, this sample composition allowed us to examine how respondents across a spectrum of expertise, from domain-specific knowledge to lay perspectives, engage with the rating framework and align with the expert-established ground truth.

\subsection{VLM Respondents}



We prompted a cohort of 19 VLMs to act as respondents in the evaluation (Table~\ref{tab:vlm_models}). These models evaluated the same audio descriptions using our multi-dimensional rating framework described in Section~\ref{sec:framework}. Each VLM was instructed to provide both ratings and accompanying justifications, allowing us to examine not only their alignment with expert-defined criteria but also how they interpret and apply the rating framework in comparison to human raters. Each model evaluated all AD versions, including outputs generated by other models as well as human-authored descriptions. This cross-evaluation design reduces the likelihood of models only “self-evaluating” and enables us to examine whether they apply the assessment criteria consistently across sources. 

\begin{table*}[htbp]
\centering
\small
\begin{tabular}{lll}
\toprule
\textbf{Source} & \textbf{Organization} & \textbf{Models} \\
\midrule

\multirow{4}{*}{Closed-source (API)}
& OpenAI 
& GPT-4o, GPT-4o mini, GPT-5, GPT-5 mini, GPT-5.2 \\

& Google (Gemini) 
& Gemini 1.5 Pro, 2.5 Flash, 2.5 Pro, 3.0 Flash, 3.0 Pro \\

& Anthropic (Claude) 
& Claude Opus 4.6, Claude Sonnet 4.6 \\

\midrule

\multirow{5}{*}{Open-source}
& Qwen 
& Qwen2.5-VL, Qwen3-VL \\

& Meta (LLaMA) 
& LLaMA 3.2 \\

& Microsoft (Phi) 
& Phi-4 \\

& Multimodal (specialized) 
& NVILA-15B, LLaVA-Video-72B, InternVL2.5-26B \\

\bottomrule
\end{tabular}
\caption{Across closed-source (API) and open-source, a total of 19 VLMs were assigned to rate ADs as respondents.}
\label{tab:vlm_models}
\Description{Table listing the 19 vision-language models used as AD raters, grouped by source. Closed-source API models include five from OpenAI (GPT-4o, GPT-4o mini, GPT-5, GPT-5 mini, GPT-5.2), five from Google Gemini (1.5 Pro, 2.5 Flash, 2.5 Pro, 3.0 Flash, 3.0 Pro), and two from Anthropic (Claude Opus 4.6, Claude Sonnet 4.6). Open-source models include Qwen 2.5 and Qwen3, LLaMA 3.2, Phi-4, and three specialized multimodal models: NVILA-15B, LLaVA-Video-72B, and InternVL2.5-26B.}
\end{table*}

\subsubsection*{Prompt Design}
Each VLM was prompted with the same rating framework given to human raters: the six dimensions, their definitions, the 1--5 scoring criteria, and examples. Alongside the video, the model received a structured JSON file containing the two inputs from Section~3.2.2: (1) the dialogue transcript extracted from the original video audio and (2) the audio description segments, each with narration text, start and end timestamps, track type (inline versus extended), and description type (visual versus text on screen). The file input contained no metadata about whether a segment originated from a human or from a particular VLM. In this setup, VLMs, like human raters, evaluated each description without knowledge of its source. VLMs rated each description, provided a justification for every score, and returned the results in a flat JSON schema. Listing~\ref{lst:prompt} shows a snippet of the prompt; the full text is provided in Appendix~\ref{appendix:prompts}.

\begin{lstlisting}[style=prompt, label={lst:prompt}, caption={A snippet of the prompt used to instruct the VLMs in applying the six-dimensional rating framework to evaluate AD.}]
PROMPT_FOR_EVALUATION = """
CONTEXT: I am providing you with two assets:
1. A video file.
2. The structured JSON data of the existing audio description.
**JSON DATA:**
```json
{json_data}
```
TASK: Analyze the video and the JSON data to evaluate the quality of the audio description track using the Multi-Dimensional Rating Framework for Audio Description.
RATING FRAMEWORK:
I. CONTENT (4 criteria based on DCMP guidelines)
II. FORMATTING (2 criteria covering how and when descriptions
are delivered)
...
OUTPUT FORMAT:
{{
"accurate_rating": "1-5",
"accurate_justification": "Justification for the rating.",
... (same rating/justification pair for prioritized, consistent,
     equal, strategic_method_selection, timing_and_placement)
}}
"""
\end{lstlisting}

\section{Modeling Rater Ability and Item Difficulty}
In Stage 5 (Fig.~\ref{fig:experimental_workflow}), we use IRT to analyze the ratings. By jointly modeling rater ability and item difficulty on a single scale, we can assess how closely each respondent aligns with our expert ground truth and how well each AD functions as a rating item.
\begingroup\let\clearpage\relax
\subsection{Partial Credit Modeling}


In the spirit of quality control, we want to be able to measure the respondents' ability to evaluate ADs relative to the experts' benchmark. To achieve this, we constructed a partial credit scoring scheme based on the distance between each respondent’s ratings and the ground truth rating for a given item by applying the Partial Credit Model (PCM) \cite{Masters1997TheModel, Masters1982AScoring}. Scores were assigned as follows:

\begin{itemize}
    \item \textbf{2 (Exact Agreement):} The respondent’s rating matched the ground truth rating.
    \item \textbf{1 (Adjacent Agreement):} The respondent’s rating differed by exactly one point from the ground truth in either direction.
    \item \textbf{0 (Distal Agreement):} The respondent’s rating differed by two or more points from the ground truth in either direction.
\end{itemize}

This recoding has certain conveniences. By having a partial credit framework, we obtain more information about a middle category (adjacent agreement) that would otherwise be lost if scored dichotomously (e.g., right vs. wrong). This allows us to evaluate rater performance by treating adjacent ratings as partial agreements, rather than dismissing them as complete errors. Therefore, each item-respondent interaction in the dataset was represented on a 0–2 scale, where 0 denotes the lowest rating alignment and 2 denotes the highest. This representation provided the foundation for the partial credit analyses presented in this section. From this, we generated person and item fit statistics along with item correlation relationships that help determine how confident we are about person and item proficiency estimates. With those estimates, we are able to plot respondents and items on the same scale visually through an Item-Respondent Map, also known as a Wright Map \cite{Masters1997TheModel}. This scale is a logarithmic scale that is interval in nature \cite{Masters1997TheModel, Masters1982AScoring}.



\subsection{Rasch Model Framework}
The PCM (a Rasch-family model) \cite{Masters1997TheModel, Masters1982AScoring} was used to estimate both item difficulty and person ability. From these, we can estimate the likelihood for particular respondents to select an item at a particular level. With three distinct rating levels, we define two cumulative Thurstonian thresholds \cite{brown2011item, burkner2019thurstonianirt} that separate each point on the scale. Threshold 1 represents the likelihood of a respondent achieving distal agreement versus adjacent and exact agreement while threshold 2 represents the likelihood of a respondent achieving exact agreement versus distal and adjacent agreement. These results are presented in the result section along with the Item-Respondent maps.




\endgroup

\section{Results}

\subsection{Fit Statistics Analysis}

We evaluated the reliability and consistency of the rating framework using Mean Square (MNSQ) fit statistics \cite{wu2013properties}. MNSQ values indicate how predictably respondents or items behave within the measurement model \cite{wilson2023constructing}. Values close to 1.00 suggest expected fit, while values substantially above 1.00 indicate greater-than-expected variability or inconsistency. In this study, we adopted the commonly accepted range of 0.75--1.33 as evidence of acceptable fit. Values above this range were interpreted as potential misfit, suggesting that a respondent or item may not have aligned consistently with the underlying evaluation construct.

We examined fit statistics at both the item and person levels. Item-level fit was used to determine whether individual evaluation items functioned consistently across respondents, while person-level fit was used to assess whether human raters and VLM respondents applied the evaluation criteria in a stable and predictable manner.

\begin{figure*}[ht]
\centering
    \includegraphics[width=\linewidth]{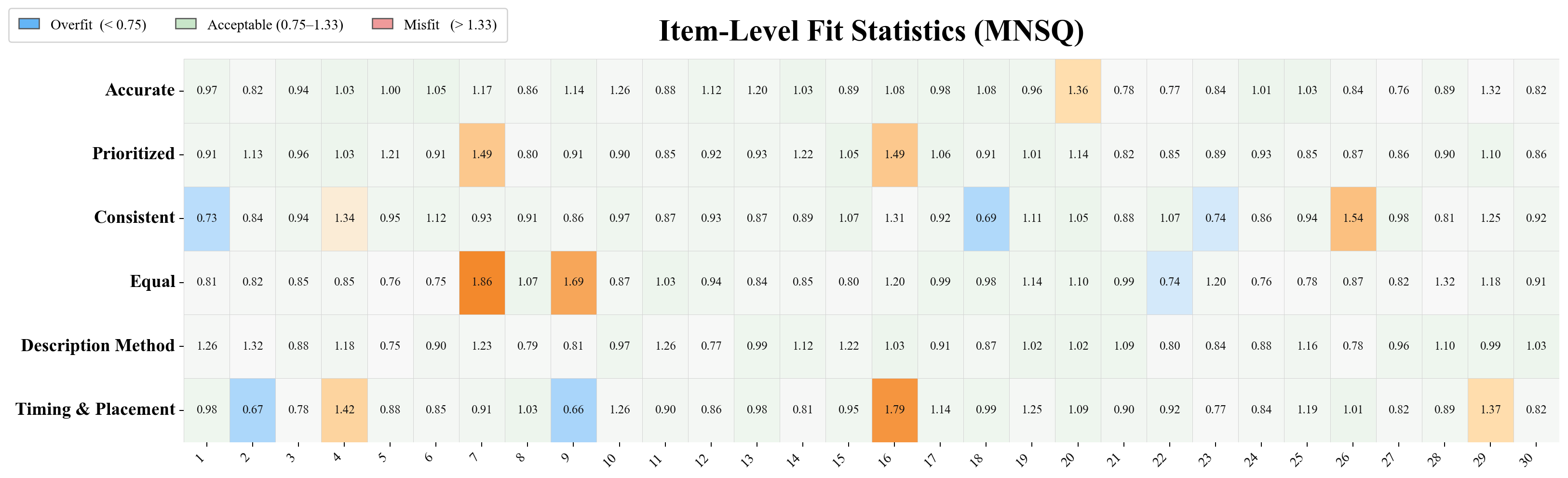}
    \Description{Heatmap of item-level infit mean-square (MNSQ) fit statistics across 30 items for the six rating dimensions: Accurate, Prioritized, Consistent, Equal, Description Method, and Timing and Placement. Values between 0.75 and 1.33 are shaded green and considered acceptable fit; values below 0.75 are shaded blue and indicate overfit; values above 1.33 are shaded orange to red and indicate misfit. Most items across all six dimensions fall within the acceptable range. Equal shows the most pronounced misfit, with item 7 reaching 1.86 and item 9 reaching 1.69, the two highest misfit values in the heatmap. Timing and Placement shows a comparable misfit at item 16 (1.79), along with a milder one at item 4 (1.42). Description Method shows no misfitting items at all, with all 30 values within the acceptable range, the cleanest fit profile among the six dimensions. Accurate, Prioritized, and Consistent show scattered isolated misfit and overfit values but are largely acceptable overall.}
    \caption{Item-level MNSQ fit statistics across the six rating dimensions.}
    \label{fig:item-fit}
\end{figure*}

\subsubsection{Item-level Reliability and Fit}

We first examined item-level fit statistics to assess whether individual evaluation items functioned consistently within the measurement model (see Table~\ref{tab:item_fit} in Appendix~\ref{Appendix:item-fit}). The majority of items fell within the acceptable MNSQ range across all dimensions, indicating that the assessment tasks were generally well aligned with the underlying construct and did not introduce substantial noise.

Only a small number of items exhibited elevated misfit. For example, Item 7 and Item 16 showed notable deviations in the Equal dimension, while Item 4 and Item 16 exceeded the acceptable range in the Timing and Placement dimension, with MNSQ values of 1.42 and 1.79, respectively. Additional isolated misfit was observed in the Prioritized dimension for Item 7 and Item 16, as well as in the Consistent dimension for Item 26. Despite these cases, item-level misfit remained limited and non-systematic. This suggests that the overall item set provided a stable and reliable basis for evaluating audio description quality across dimensions.

\begin{figure*}[ht]
\centering
    \includegraphics[width=\linewidth]{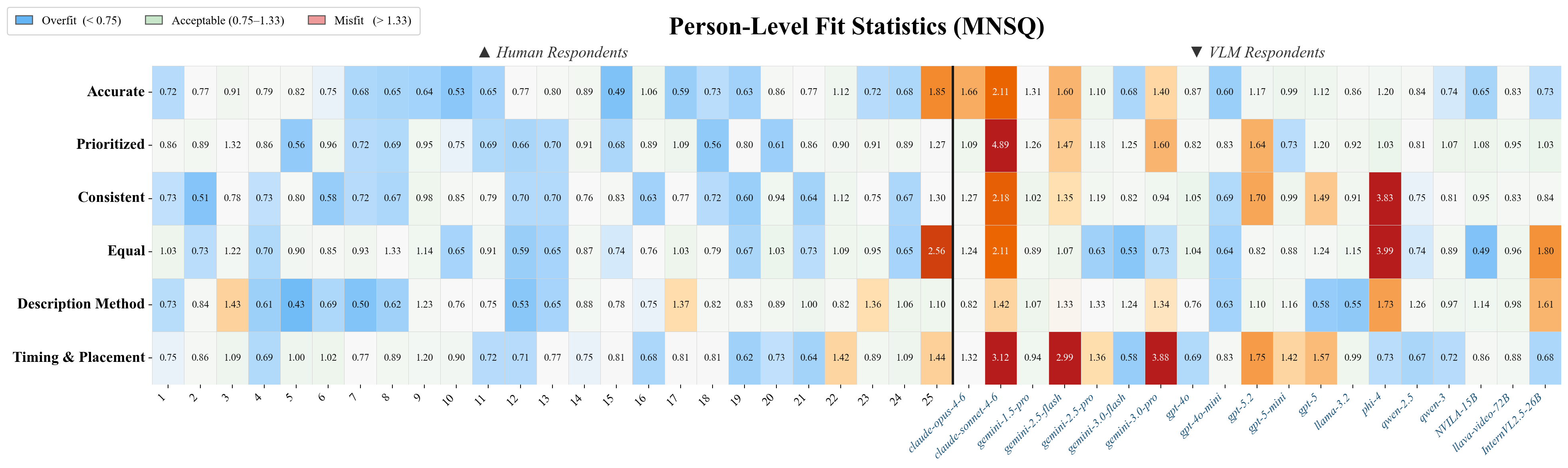}
    \Description{Heatmap of person-level infit mean-square (MNSQ) fit statistics for 44 respondents, 25 human raters numbered 1 through 25 and 19 VLMs listed by model name, across the six rating dimensions. A vertical line separates human respondents on the left from VLM respondents on the right. Most human respondents show overfit or acceptable values, indicating consistent rating patterns, with one notable exception: human respondent 25 shows elevated misfit on multiple dimensions, including Equal (2.56) and Accurate (1.85). Fit is far more variable among VLMs. Claude Sonnet 4.6 shows severe misfit across nearly all dimensions, most extremely on Prioritized (4.89), the highest value in the entire heatmap, along with Consistent (2.18), Accurate (2.11), and Equal (2.11). Phi-4 shows severe misfit concentrated on Consistent (3.83) and Equal (3.99). Gemini 3.0 Pro shows severe misfit isolated to Timing and Placement (3.88) while showing overfit on most other dimensions. Gemini 2.5 Flash also shows elevated misfit on Timing and Placement (2.99). Other VLMs, including GPT-4o, GPT-4o mini, Llama 3.2, and NVILA-15B, show fit values comparable to the acceptable range seen in most human respondents.}
    \caption{Person-level MNSQ fit statistics for human and VLM respondents across the six rating dimensions.}
    \label{fig:person-fit}
\end{figure*}

\subsubsection{Person-Level Reliability and Fit}

We next assessed respondent-level consistency using person-level MNSQ fit statistics (see Table~\ref{tab:fit_statistics} in Appendix~\ref{Appendix:person_fit_stats}). Overall, many respondents, including both human raters and VLMs, fell within the acceptable range across dimensions. This suggests that the rating framework was generally applied in a consistent and predictable manner.

Among human raters, misfit was relatively limited and sporadic. However, a small number of individuals exhibited elevated misfit on specific dimensions. In particular, Human25 showed substantial deviations across multiple dimensions, including Accurate = 1.85, Equal = 2.56, and Timing and Placement = 1.44. This pattern suggests inconsistent interpretation or application of the evaluation criteria. Other instances of human-rater misfit were isolated and did not indicate a broader pattern across the cohort.

In contrast, misfit among VLM respondents was more systematic and model-dependent. Certain models, such as Claude Sonnet 4.6 and Gemini 2.5 Flash, exhibited consistently high misfit across multiple dimensions, suggesting unstable or overly variable rating behavior. Other models, including GPT-4o mini and Qwen 2.5, remained largely within acceptable bounds, demonstrating more stable alignment with the evaluation framework. These results indicate that while VLMs can approximate consistent rating behavior, their reliability varies substantially across models. This variation should be considered when interpreting the Item-Respondent Maps and comparing VLM performance against human evaluation patterns.

\subsection{Item-Respondent Map Interpretation}
To better understand how both respondents and items align across the latent or logit scale, we examined Item-Respondent maps for each of the six dimensions. These visualizations place a total of 44 respondents (humans and VLMs) on the left side as distribution plots according to their proficiency estimates ($\theta$), logits being in the middle (y-axis), and items with their thresholds on the right, providing a direct comparison of rater ability and item difficulty. Items were ordered by their threshold 2 logits in increasing order. Beyond targeting, we also examine the banding of each dimension: how cleanly the item thresholds function as waypoints, along the latent scale that separate respondents into distinguishable regions of ability. Clear banding means human and VLM ability estimates separate distinctly once a waypoint is drawn; fuzzy banding means the estimates cluster around the waypoint itself, so the waypoint does not reliably distinguish higher from lower performing raters.

\subsubsection{Content Dimensions Analysis}

\begin{figure*}[h]
\centering
\begin{subfigure}{0.48\textwidth}
    \centering
    \includegraphics[width=\linewidth]{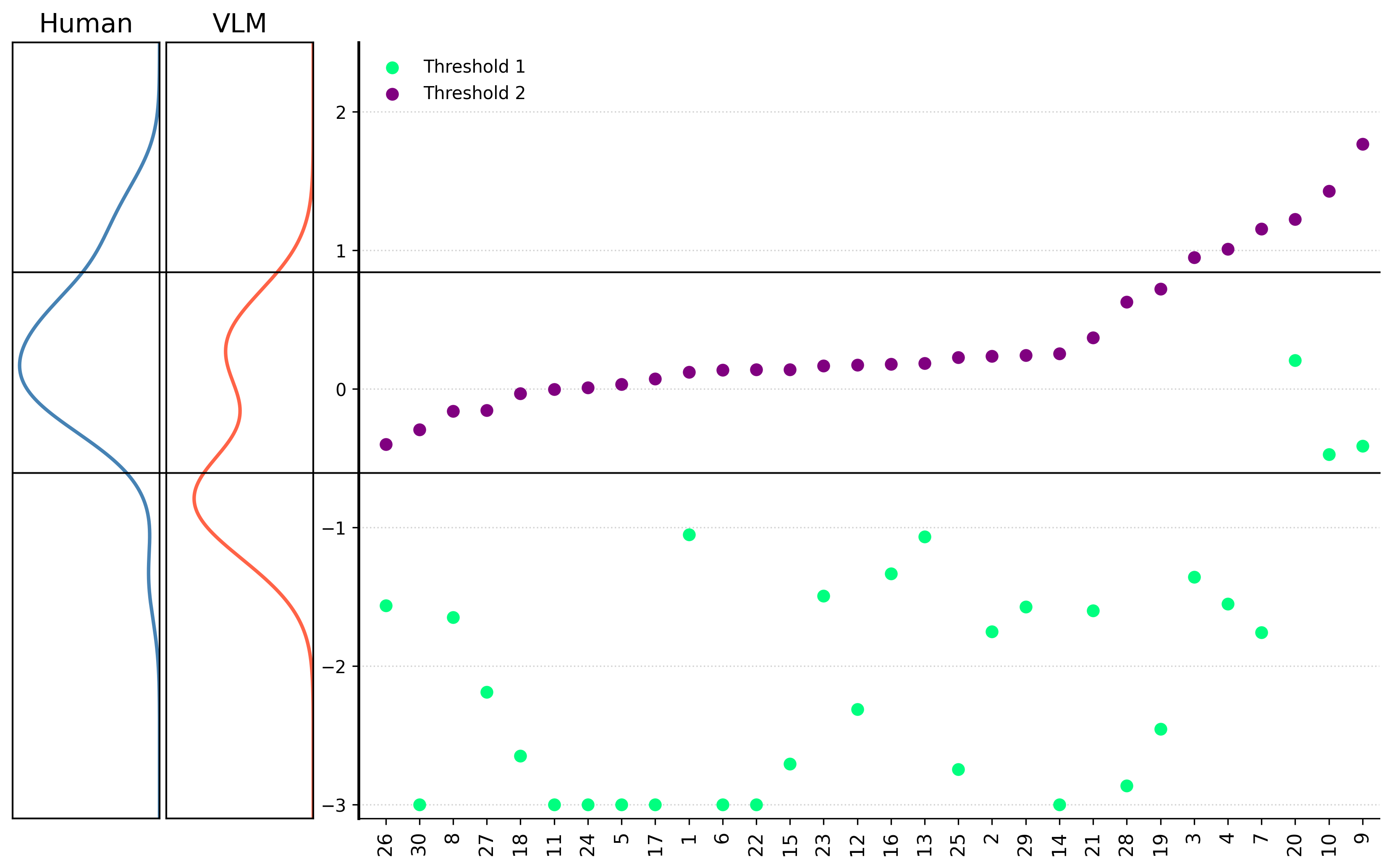}
    \Description{Item-respondent map for the Accurate dimension. Human ability estimates form a single narrow peak around 0.3 to 0.5 logits. VLM ability estimates show a bimodal distribution, with a main peak similar to the human peak and a small secondary bump extending above 1 logit. Thirty items are plotted at right, ordered by threshold 2 difficulty. Threshold 2 values rise gradually from about -0.5 to 1.8 logits, with most items clustered below 0.5 and a handful of items above 1. Threshold 1 values remain mostly between -3 and -1 logits, except for three items (20, 10, 9) whose threshold 1 sits notably higher, near 0. Human and VLM ability estimates separate reasonably clearly around the upper waypoint, indicating clear banding for this dimension.}
    \caption{Accurate}
    \label{fig:accurate}
\end{subfigure}
\hfill
\begin{subfigure}{0.48\textwidth}
    \centering
    \includegraphics[width=\linewidth]{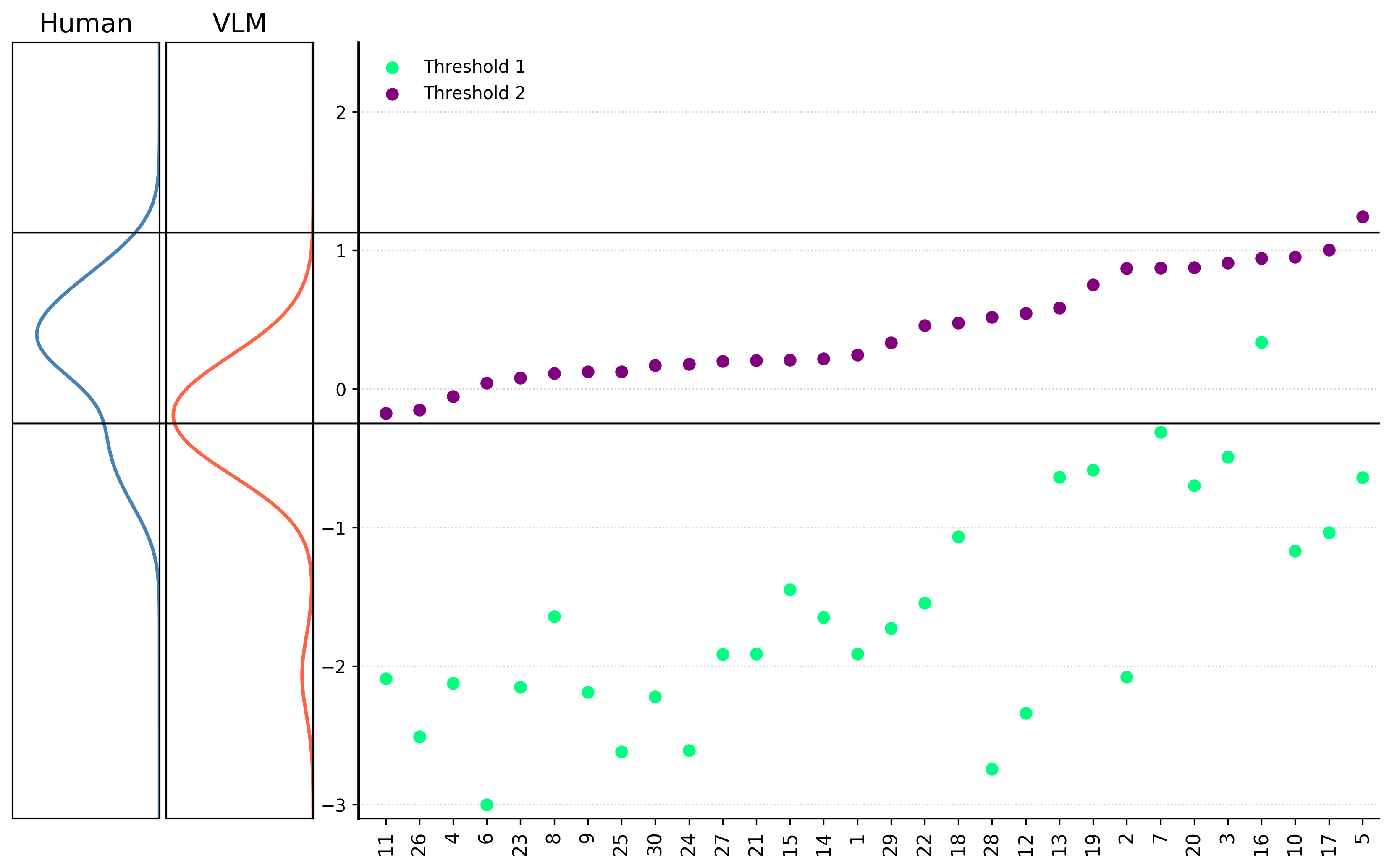}
    \Description{Item-respondent map for the Prioritized dimension. Human ability estimates form a narrow peak around 0.3 to 0.5 logits. VLM ability estimates peak slightly lower, closer to 0, with a broader spread below that peak. Thirty items are plotted at right, ordered by threshold 2 difficulty. Threshold 2 values rise smoothly from about -0.2 to 1.25 logits, a narrower range than Accurate. Threshold 1 values remain mostly between -3 and -1 logits, except for one item (16) whose threshold 1 sits notably higher, near 0.4. Banding around the upper waypoint remains reasonably clear, though less pronounced than in Accurate or Consistent.}
    \caption{Prioritized}
    \label{fig:prioritized}
\end{subfigure}
\vspace{0.5em}
\begin{subfigure}{0.48\textwidth}
    \centering
    \includegraphics[width=\linewidth]{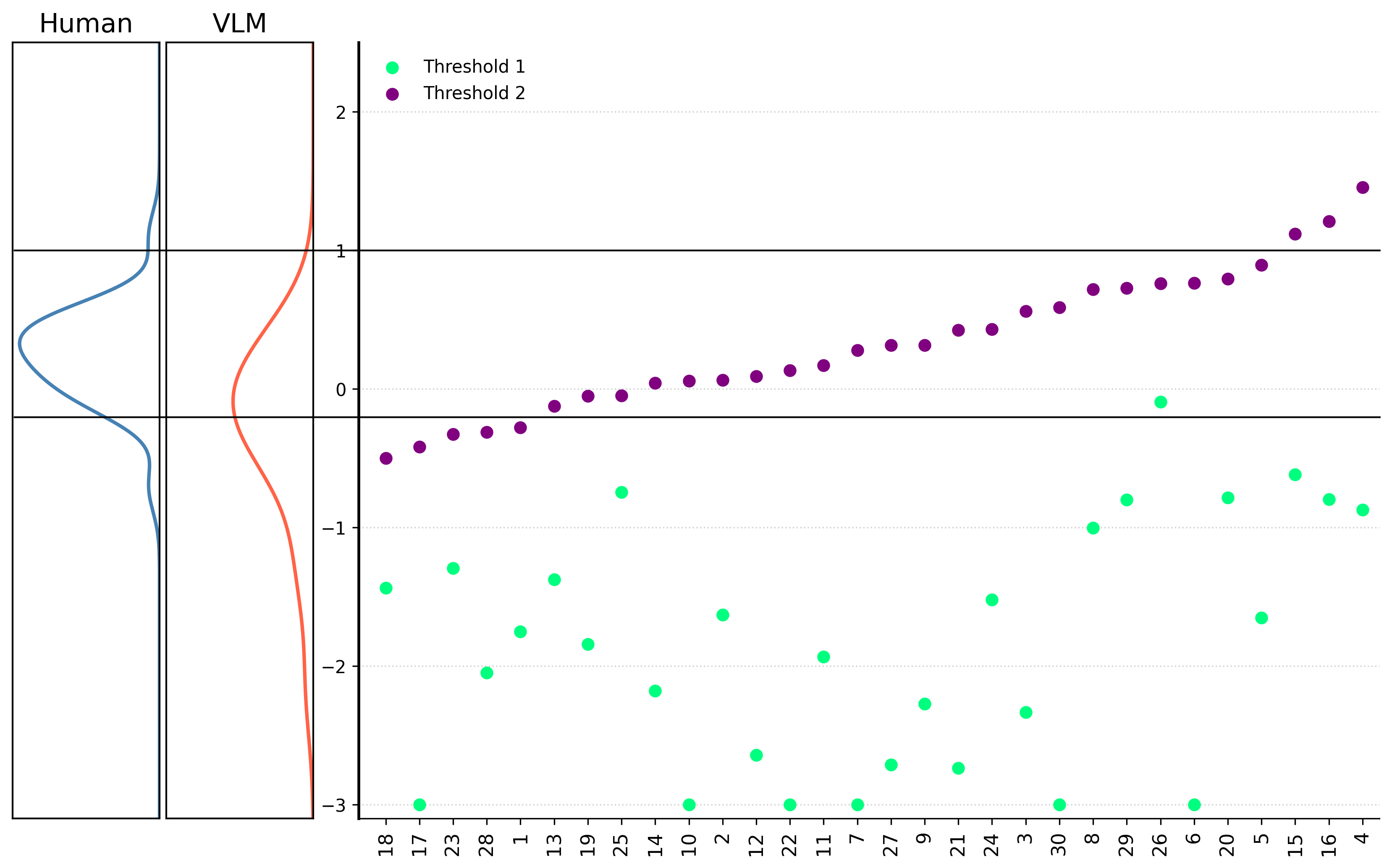}
    \Description{Item-respondent map for the Consistent dimension. Human ability estimates form a narrow peak around 0.3 to 0.5 logits. VLM ability estimates show a similar main peak with a small secondary bump extending down toward -1.5, giving a slightly bimodal shape. Thirty items are plotted at right, ordered by threshold 2 difficulty. Threshold 2 values rise steadily from about -0.5 to 1.75 logits, with the final three items jumping noticeably higher than the rest. Threshold 1 values remain mostly between -3 and -0.7 logits, except for one item (20) whose threshold 1 sits notably higher, near 0.3. Human and VLM ability estimates separate distinctly around the upper waypoint, indicating clear banding for this dimension.}
    \caption{Consistent}
    \label{fig:consistent}
\end{subfigure}
\hfill
\begin{subfigure}{0.48\textwidth}
    \centering
    \includegraphics[width=\linewidth]{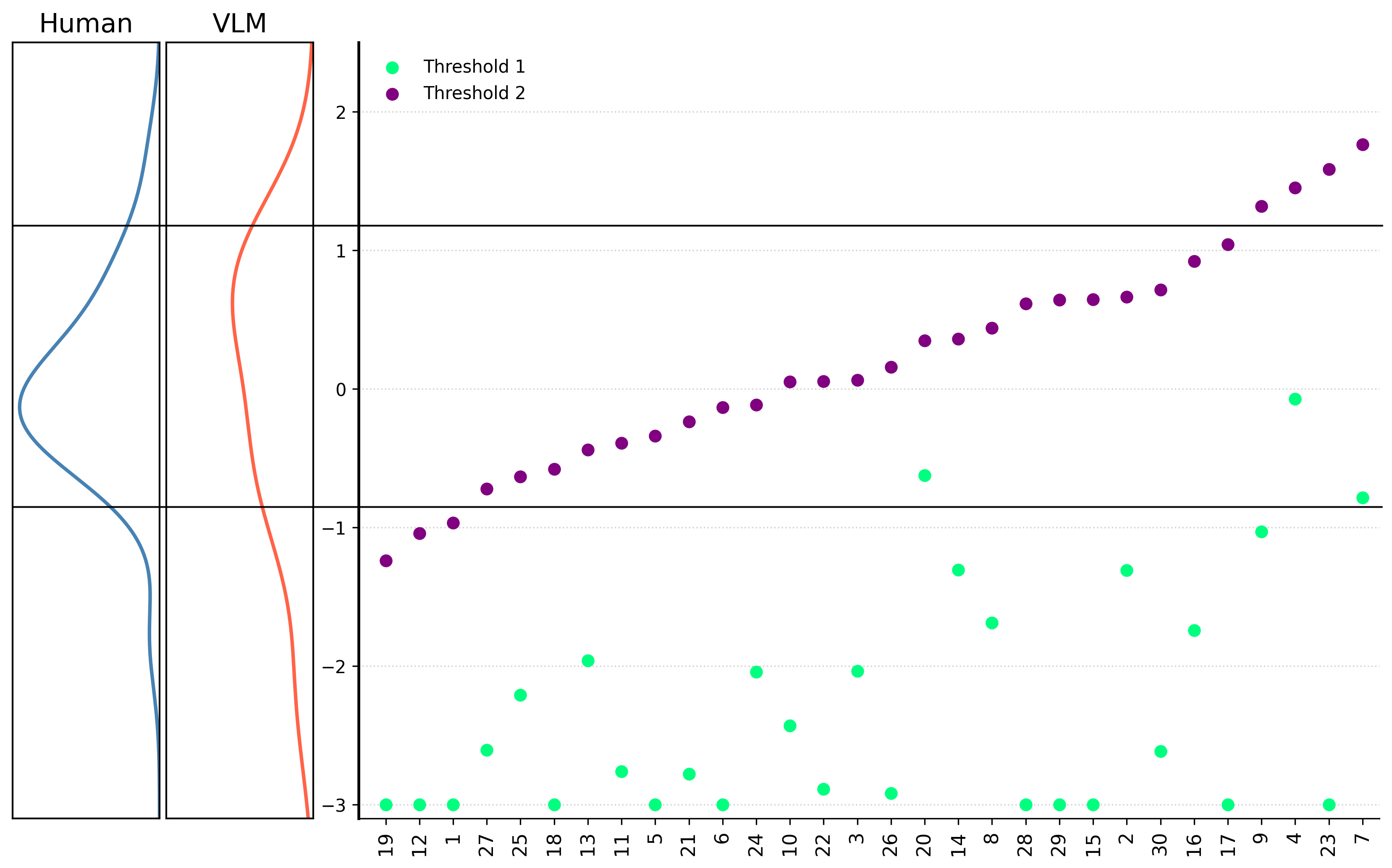}
    \Description{Item-respondent map for the Equal dimension. Human ability estimates form a narrow peak around 0.3 to 0.5 logits. VLM ability estimates peak lower, around 0, with a broader spread and a long tail extending down toward -3. Thirty items are plotted at right, ordered by threshold 2 difficulty. Threshold 2 values rise gradually from about -1.3 to roughly 0.5 logits before jumping sharply for the last several items, reaching 1.75 logits at the most difficult item, the widest threshold 2 range among the six dimensions. Threshold 1 values are scattered widely, with several items in the -3 to -1 range and several others elevated closer to 0. Human and VLM ability estimates converge closely around the waypoint rather than separating cleanly, indicating the fuzziest banding among the six dimensions.}
    \caption{Equal}
    \label{fig:equal}
\end{subfigure}
\caption{Item-Respondent Maps for content-related dimensions.}
\Description{Four item-respondent maps (Wright maps) for the content-related rating dimensions: Accurate, Prioritized, Consistent, and Equal. Each map shows human and VLM rater ability distributions on the left and item threshold difficulties on the right, with threshold 1 (lower) in green and threshold 2 (upper) in purple. Across all four dimensions, threshold 1 values remain well below rater abilities while threshold 2 values overlap with the rater distribution, indicating that items are easy to pass at the lower rating level but discriminate meaningfully at the higher level.}
\label{fig:content_dimensions}
\end{figure*}

Across the four content-related dimensions in Figure~\ref{fig:content_dimensions}, the Item Respondent maps show generally well-structured measurement spaces, with item thresholds spanning a broad range of difficulty and respondent abilities are concentrated within a narrower band near average or slightly above. In the Accurate dimension (Fig. \ref{fig:accurate}) and the Consistent dimension (Fig. \ref{fig:consistent}), item thresholds (particularly the second thresholds) are distributed from near 0 to above 1.5 logits, while most respondents, both human and VLM, cluster between approximately -1 and 1 logits. This indicates moderate to good targeting, although the upper range of item difficulty slightly exceeds the ability of most respondents, suggesting that some items with highly difficult distinctions are less reliably evaluated. These two dimensions also show the clear banding: human and VLM ability estimates separate distinctly around the waypoint, indicating the instrument reliably distinguishes competent from unreliable raters.

In contrast, Prioritized (Fig. \ref{fig:prioritized}) and Equal (Fig. \ref{fig:equal}) dimensions exhibit greater variability and less stable alignment. While the Prioritized dimension (Fig. \ref{fig:prioritized}) maintains a relatively smooth progression of item difficulty, respondent distributions are more compressed, indicating limited differentiation among raters. Banding around the waypoint in Prioritized nonetheless remains reasonably clear, though less pronounced than in Accurate or Consistent. The Equal dimension (Fig. \ref{fig:equal}) shows the greatest dispersion in both item thresholds and respondent abilities, with several respondents occupying higher logit regions and items extending beyond 1.5 logits. This wider spread suggests that Equal (Fig. \ref{fig:equal}) captures more subjective or interpretive aspects of evaluation, leading to increased variability in both human and VLM responses. Correspondingly, Equal shows the fuzziest banding of the four dimensions: human and VLM ability estimates converge closely around the waypoint rather than separating cleanly. Notably, VLM distributions tend to be slightly shifted toward lower ability regions compared to humans in these dimensions except for the Equal dimension.

\subsubsection{Formatting Dimensions Analysis}
\begin{figure*}[htpb]
\centering
\begin{subfigure}{0.45\textwidth}
    \centering
    \includegraphics[width=\linewidth]{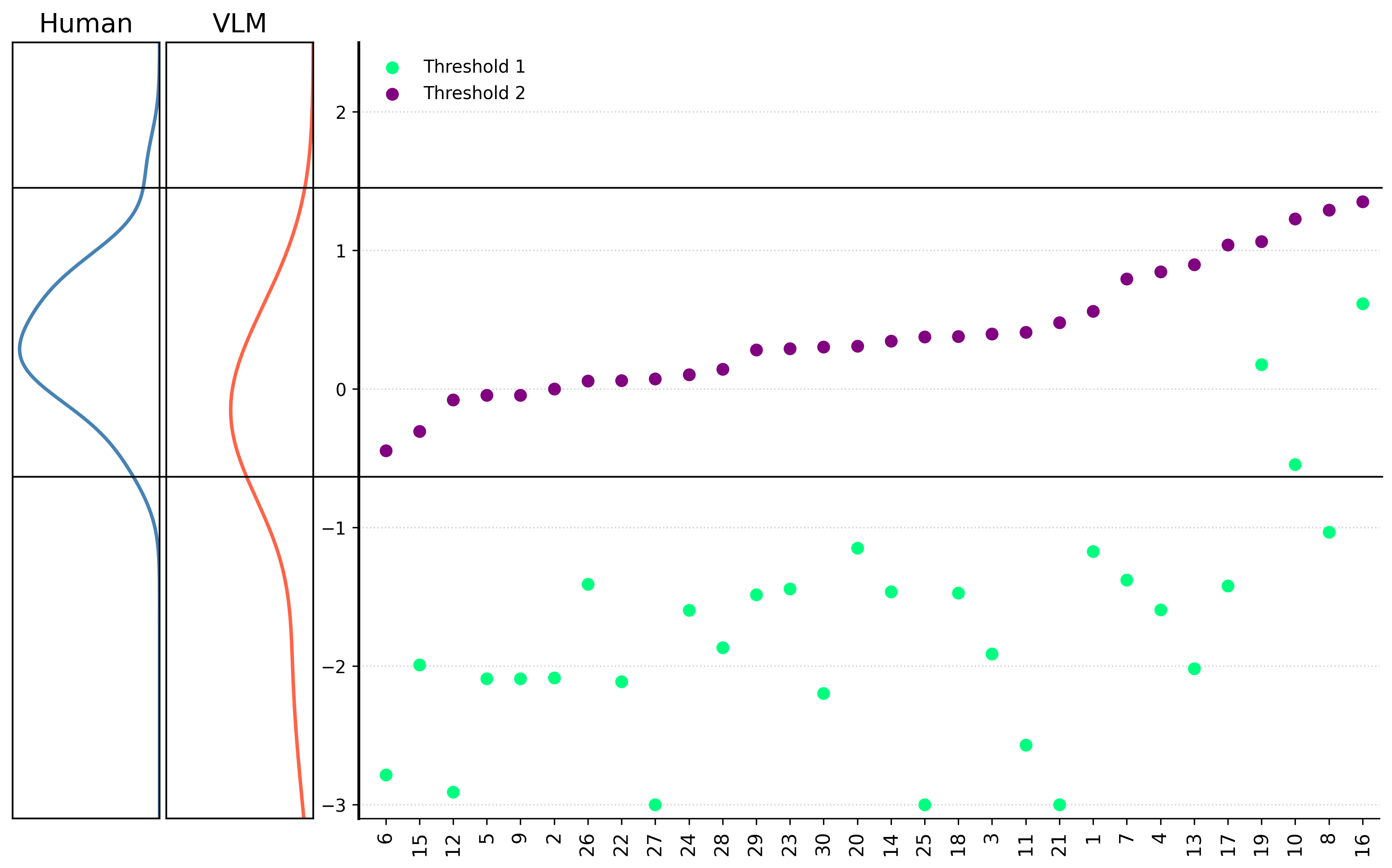}
    \Description{Item-respondent map for the Description Method dimension. Human ability estimates form a narrow peak around 0.3 to 0.5 logits. VLM ability estimates peak lower, around 0, with a broad, elongated spread and a long tail extending down toward -3. Thirty items are plotted at right, ordered by threshold 2 difficulty. Threshold 2 values rise smoothly from about -0.6 to 1.4 logits, reaching the highest difficulty levels among all six dimensions. Threshold 1 values remain mostly between -3 and -1 logits, except for one item (16) whose threshold 1 sits well above the rest, near 0.6. A large share of VLM ability estimates fall below the range distinguishable by the waypoints, indicating this dimension is limited primarily by targeting rather than banding.}
    \caption{Description Method}
    \label{strategy}
\end{subfigure}
\hfill
\begin{subfigure}{0.45\textwidth}
    \centering
    \includegraphics[width=\linewidth]{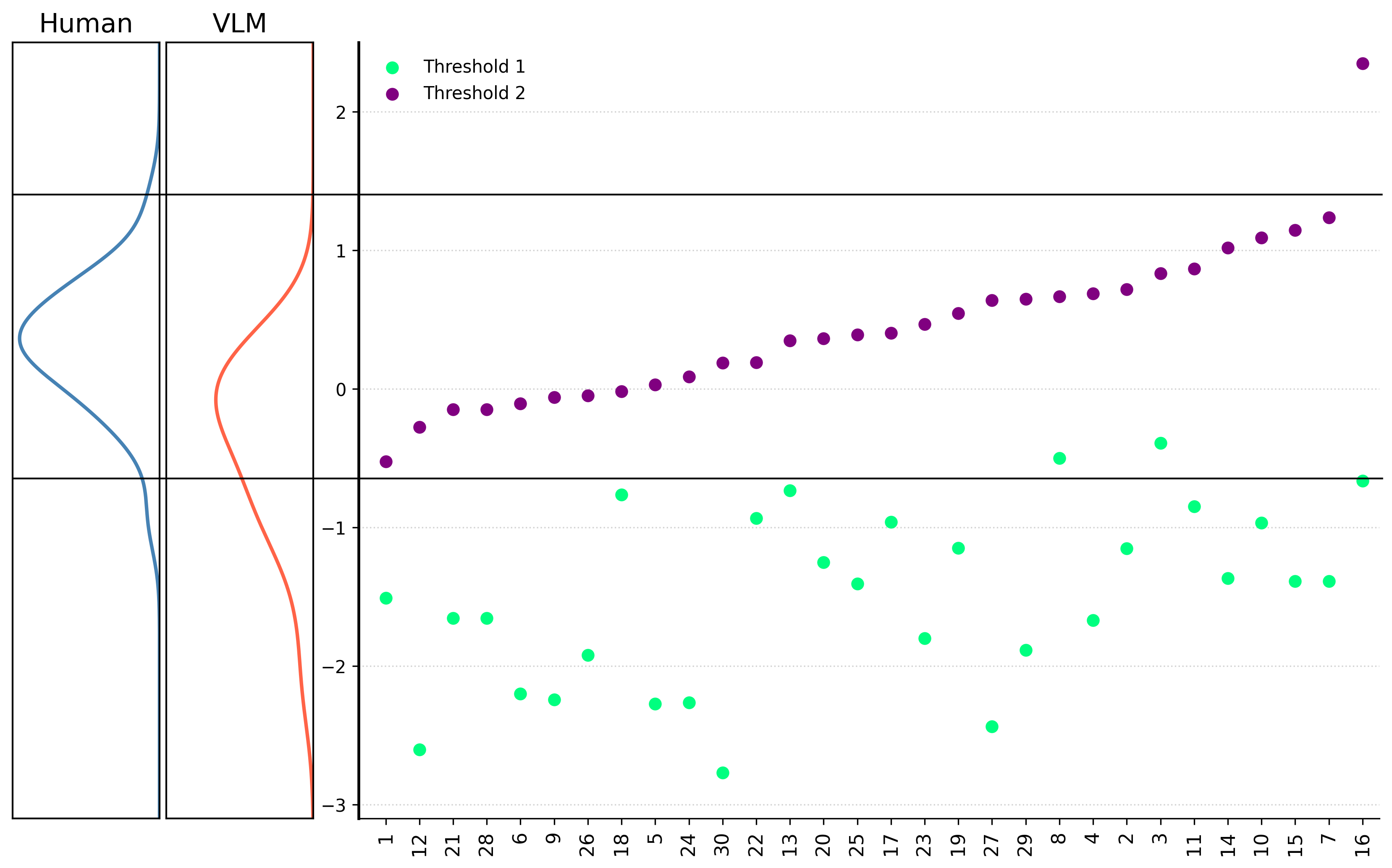}
    \Description{Item-respondent map for the Timing and Placement dimension. Human ability estimates form a narrow peak around 0.3 to 0.5 logits. VLM ability estimates show a bimodal distribution, with a main peak similar to the human peak and a secondary bump extending above 1 logit. Thirty items are plotted at right, ordered by threshold 2 difficulty. Threshold 2 values rise gradually from about -0.7 to 1.3 logits before jumping sharply to 2.4 logits at a single, isolated item (16). Threshold 1 values remain mostly between -3 and -1 logits, with several items elevated between -1.5 and 0. Excluding the isolated high-difficulty item, separation around the waypoints remains reasonably clear.}
    \caption{Timing \& Placement}
    \label{timing}
\end{subfigure}
\hfill
\caption{Item-Respondent Maps for formatting-related dimensions.}
\label{formatting_dimensions}
\Description{Two item-respondent maps (Wright maps) for the formatting-related rating dimensions: Description Method and Timing and Placement. Each map shows human and VLM rater ability distributions on the left and item threshold difficulties on the right, with threshold 1 (lower) in green and threshold 2 (upper) in purple. In both dimensions, threshold 1 values sit well below rater abilities while threshold 2 values overlap with the rater distribution, consistent with the pattern observed in the content-related dimensions.}
\end{figure*}

The formatting-related dimensions (Figure \ref{formatting_dimensions}) exhibit a distinct structural pattern compared to content dimensions. Respondent abilities, especially among VLMs, are more concentrated in average to lower logit ranges, resulting in noticeable gaps between item difficulty and respondent ability. This suggests that these dimensions impose greater cognitive or interpretive demands, making them more challenging for both human and model-based evaluators. This is primarily a targeting focus rather than a banding one where respondent ability does fall within the measurable range, waypoint separation remains reasonably clear, but a large share of VLM ability estimates fall below where the instrument can distinguish performance at all.

Additionally, the Description Method dimension demonstrates the most pronounced separation between human and VLM distributions, with VLMs more frequently occupying lower ability regions. This pattern indicates that VLMs struggle more with higher-level evaluative reasoning required in strategy selection, compared to more surface-level judgments. The Timing \& Placement dimension (Fig. \ref{timing}), while showing a similar trend, exhibits slightly better alignment, suggesting it is comparatively easier to evaluate. Timing \& Placement also shows clearer banding within its measurable range than Description Method, aside from a single unusually difficult item that extends its threshold spread. Overall, the Item Respondent maps indicate that evaluators handle structural versus interpretive aspects of audio description quality differently on content-related and formatting-related dimensions. This pattern is also reflected at the model level, where only a small subset of VLMs (e.g., GPT-4o mini and Qwen 2.5) approach the central ability range in the Timing \& Placement dimension (Fig. \ref{timing}), while most models remain in lower or random logit regions, particularly in the Description Method dimension (Fig. \ref{strategy}).

\subsection{Observations on VLM Reasoning}

While VLMs approximated ground-truth ratings to a degree comparable with human raters across most dimensions, their justifications showed limited grasp of the assessment framework and rarely offered actionable feedback for improvement. By contrast, human raters provided precise, scene-specific comments that are valuable even when their numeric ratings diverged from the expert-defined ground truth. Table~\ref{tab:qualitative} presents representative items for each of the six dimensions, drawn from all three genres and from both human-authored and VLM-generated descriptions, pairing ground-truth, human, and VLM ratings with their rationales. Human rationales were optional for our raters given time constraints and are therefore not available for every rating; the items shown were necessarily drawn from those for which a rationale was recorded.

\begin{table*}[htbp]
\centering
\small
\setlength{\tabcolsep}{3.5pt}
\renewcommand{\arraystretch}{1.15}
\caption{One representative item per dimension of the rating framework, drawn from all three genres and from both human-authored and VLM-generated descriptions. $|\Delta|$ is the absolute difference between a rating and the ground truth (GT). Items were selected to illustrate justification quality; the deltas shown are not representative of overall rating agreement. Human justifications were optional; items were drawn from ratings for which one was recorded. Justifications are verbatim, trimmed with [\ldots] where noted.}
\Description{Table with one example item per rating dimension. Each row group names the dimension, video, AD source, and ground truth rating, then pairs one human rater and one VLM rater with their ratings, deltas from ground truth, and verbatim justifications. Accurate: Pickles video, Qwen2.5-VL AD, ground truth 4; Human R6 rated 4, asking where the canning funnel and brine descriptions are; Phi-4 rated 4, stating descriptions were generally correct but had noticeable errors without naming any. Prioritized: Star Wars, GPT-4o AD, ground truth 5; Human R3 rated 4, suggesting Leia should have been named instead of a woman with braided hair; LLaMA 3.2 rated 4, stating it prioritizes important elements but sometimes focuses on minor details. Consistent: Origami, Qwen2.5-VL AD, ground truth 3; Human R6 rated 1, citing inconsistent shape terms such as slight curve versus symmetric kite shape for the same folds; Gemini 3.0 Flash rated 4, stating terminology is stable with a minor inconsistency. Equal: Elf, volunteer AD, ground truth 3; Human R3 rated 2, citing interpretation such as looks extremely shocked and explaining flashbacks rather than describing them; NVILA-15B rated 3, stating the description is generally objective. Strategic Use of Description Method: Crash Course Kid, volunteer AD, ground truth 3; Human R7 rated 2, citing irrelevant extended clips such as reading Big Question aloud when the speaker already said it; Qwen2.5-VL rated 3, stating extended descriptions are sometimes used where inline would suffice. Timing and Placement: Star Wars, GPT-4o AD, ground truth 4; Human R3 rated 4, noting her stance is described slightly too early before the TIE fighter approaches; GPT-4o-mini rated 4, stating descriptions are generally well-timed with occasional slight mistiming. Across rows, human justifications cite specific scenes and moments while VLM justifications remain general.}
\label{tab:qualitative}
\begin{tabularx}{\textwidth}{@{}>{\raggedright\arraybackslash}p{1.95cm} p{2.0cm} p{1.5cm} c p{1.9cm} cc X@{}}
\toprule
\textbf{Dimension} & \textbf{Video} & \textbf{AD Source} & \textbf{GT} &
\textbf{Rater} & \textbf{Rating} & \textbf{$|\Delta|$} & \textbf{Justification} \\
\midrule

Accurate & \textit{Pickles} \newline (How-to) & Qwen2.5-VL & 4
& Human (R6) & 4 & \textbf{0} & Description didn't note to use a canning funnel. Where is the description of brine for each recipe? \\
& & & & Phi-4 & 4 & \textbf{0} & The descriptions were generally correct but had noticeable errors. Some details were missing or included too many unnecessary details. \\
\addlinespace[2pt] \cmidrule(l){1-8} \addlinespace[2pt]

Prioritized & \textit{Star Wars} \newline (Entertain\-ment) & GPT-4o & 5
& Human (R3) & 4 & \textbf{1} & Provided a good description of a woman with braided hair, but perhaps they should have said Leia by name. \\
& & & & LLaMA 3.2 & 4 & \textbf{1} & Prioritizes the most important visual elements, but sometimes focuses too much on minor details. \\
\addlinespace[2pt] \cmidrule(l){1-8} \addlinespace[2pt]

Consistent & \textit{Origami} \newline (How-to) & Qwen2.5-VL & 3
& Human (R6) & 1 & \textbf{2} & Inconsistent shape terms, `slight curve' for what are only angles (a curve is rounded), vs.\ the precise `symmetric kite shape' elsewhere; the vocabulary shifts between exact and contradictory for the same folds. \\
& & & & Gemini 3.0 Flash & 4 & \textbf{1} & Terminology is stable throughout, though there is a minor inconsistency. \\
\addlinespace[2pt] \cmidrule(l){1-8} \addlinespace[2pt]

Equal & \textit{Elf} \newline (Entertain\-ment) & Volunteer & 3
& Human (R3) & 2 & \textbf{1} & Too much interpretation, such as `looks extremely shocked,' and tells why he is having flashbacks instead of describing the flashback scenes. \\
& & & & NVILA-15B & 3 & \textbf{0} & The audio description is generally objective with rare minor interpretive moments.\\

\addlinespace[2pt] \cmidrule(l){1-8} \addlinespace[2pt]

Strategic Use of Description Method & \textit{Crash Course Kid} \newline (Education) & Volunteer & 3
& Human (R7) & 2 & \textbf{1} & Very little description as not much is needed, however poor choices when descriptions were used. Extended clips were irrelevant, such as saying `Big Question,' when the speaker noted she would answer a big question. [\ldots] \\
& & & & Qwen2.5-VL & 3 & \textbf{0} & The use of inline and extended descriptions is somewhat appropriate, but there are instances where extended descriptions are used when inline might have been sufficient. \\
\addlinespace[2pt] \cmidrule(l){1-8} \addlinespace[2pt]

Timing and Placement & \textit{Star Wars} \newline (Entertain\-ment) & GPT-4o & 4
& Human (R3) & 4 & \textbf{0} & Describes her stance slightly too early before the TIE fighter approaches. \\
& & & & GPT-4o-mini & 4 & \textbf{0} & Descriptions are generally well-timed but occasionally experience slight mistiming or overlap with critical audio. \\

\bottomrule
\end{tabularx}
\end{table*}

VLM explanations frequently sounded plausible but misapplied evaluation criteria. For example, Gemini 1.5 Pro argued that certain extended descriptions could be inline because \textit{``the dialogue has pauses,''} while GPT-4o claimed a description could be shortened \textit{``without losing detail.''} In reality, these clips contained no audio gaps, making inline delivery impossible. Such reasoning reflects a surface-level application of the evaluation criteria without a grounded understanding of delivery constraints. More often, VLMs offered vague comments with little diagnostic value. For instance, GPT-4o-mini justified a rating of 4 on the Timing and Placement dimension by simply stating that descriptions were \textit{``generally well-timed but occasionally experience slight mistiming,''} without identifying where the problem occurred. Similarly, a rating of 4 on the Accurate dimension from Phi-4 was justified with \textit{``The descriptions were generally correct but had noticeable errors. Some details were missing\ldots''},  but the VLM gave no concrete example of these \textit{``details''}.

Human raters, in contrast, anchored their feedback in specific scenes and provided constructive suggestions. When giving a description a lower score on Timing and Placement, a human rater noted, \textit{``describes her stance slightly too early before the TIE fighter approaches.''} When giving a lower score on Prioritized, the rater wrote, \textit{``provided a good description of the woman with braided hair, but perhaps they should have said Leia by name.''} Such feedback directly identifies what went wrong and how it could be improved.

The contrast sharpens further when comparing VLM reasoning to the formative feedback from our blind QC specialist. Reviewing a number of human-authored and AI-generated descriptions in her own professional QC format, independent of our framework, she produced observations that mapped directly onto our framework's dimensions. For example, in the Bald Eagle video, she flagged that the description \textit{``eventually loses track of which eagle is which and refers to both simply as a bald eagle,''} leaving the audience confused about the outcome, an issue that corresponds to the Consistent dimension. On the Star Wars trailer, she noted that the video \textit{``pauses at several points to accommodate descriptive details''} though \textit{``there may be sufficient natural space in the original audio track to avoid pausing the video,''} suggesting extended narration was used where inline would have sufficed, a Strategic Use of Description Method issue. She also flagged a timing issue, noting that the line \textit{``She looks at the camera and winks''} was slightly off. These observations exemplify what dimension-specific, scene-anchored diagnostic feedback looks like in professional QC practice. The framework was designed to capture this kind of feedback; VLM justifications, despite their numerical alignment with expert ratings, do not yet approximate it.

These findings highlight an important distinction. VLM ratings can approximate the benchmark across dimensions, but their justifications currently lack the specificity needed to guide system improvement. Human feedback generates actionable insights that can improve volunteer-authored and AI-generated AD. In this way, VLMs can contribute scalable ratings, but human input is essential for ensuring that evaluation leads to better practice and system development.

\section{Discussion}

Our study applied a methodological workflow, grounded in Item Response Theory, to evaluate whether VLMs and humans can reliably assess audio description quality on full-length video. The findings reveal both promise and cautions for scalable AD evaluation, with implications for how the accessibility community integrates AI into quality control systems.

\subsection{Humans and VLMs Align Comparably, But Not Identically}
Across all six dimensions, human and VLM raters aligned with expert ground truth to comparable degrees. Humans slightly outperformed VLMs on most dimensions, but the differences were small, and the ability distributions of the two groups substantially overlapped across all dimensions (Fig.~\ref{fig:content_dimensions} and Fig.~\ref{formatting_dimensions}). Despite differences in how humans and VLMs approach subjective rating tasks, our results indicate that neither group consistently outperformed the other in alignment with expert consensus. For accessibility practitioners and AD platforms considering VLMs as raters, this suggests that current VLMs can approximate human performance at the score level.

However, score-level alignment captures only part of evaluation quality. The Item Respondent map reports how closely a rater's responses match the ground-truth ratings established by expert consensus (Section~\ref{sec:procedure}), aggregated across items. A rater who aligns closely with ground truth may do so because they have calibrated expertise, because they hedge toward the middle of the scale, because they benefit from group-level averaging, or because they pattern-match without deep understanding. These are very different behaviors that look identical on an alignment metric. Our qualitative analysis reveals this limitation: VLM justifications were often generic or misapplied evaluation criteria, whereas human raters, even when achieving lower numerical alignment, more frequently identified specific issues and provided actionable feedback (Table~\ref{tab:qualitative}). Thus, evaluating VLM raters should consider not only score alignment but also whether their rationales are grounded, specific, and diagnostically useful.

Our current prompt requested justifications but did not require supporting evidence, making explanations difficult to verify. Since our results identify unreliable justification as a VLM failure mode, future iterations should explicitly prompt for evidence such as timestamps or scene references that can be cross-checked against the AD track and dialogue transcript. A natural extension of our workflow is to incorporate justification quality directly into the scoring model, so that raters whose justifications are incorrect or generic weigh less in the model than those whose justifications offer high diagnostic utility.

\subsection{Task-Specific Evaluation Is Necessary for VLM Raters}

\begin{figure}[htbp]
\centering
\includegraphics[width=\linewidth]{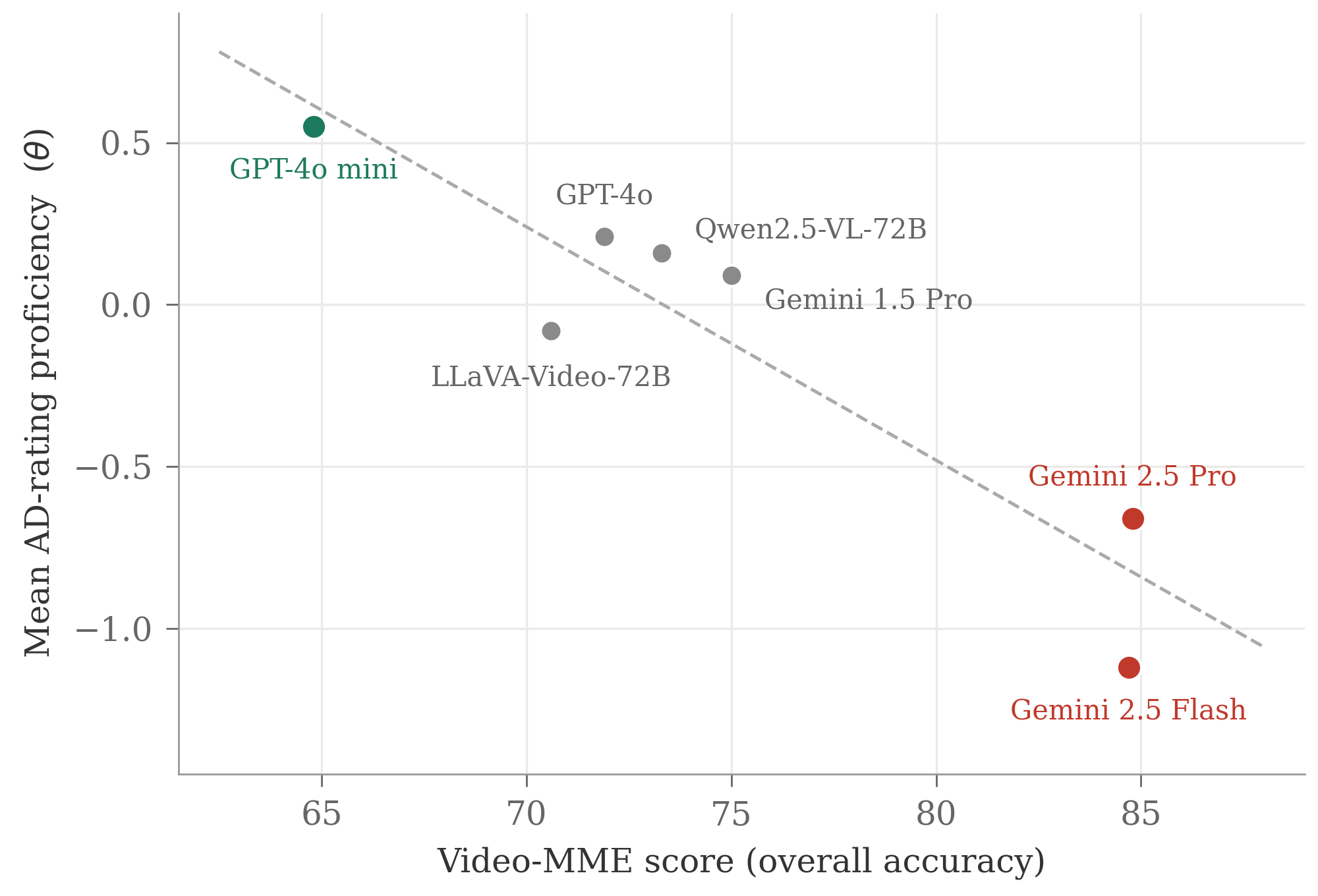}
\caption{Video-MME score against mean AD-rating proficiency ($\theta$) for the seven models with reported benchmark scores. The two highest scorers on Video-MME rank lowest as AD raters; the lowest scorer rates best. Gemini 2.5 scores are developer-reported rather than from the public leaderboard. Dashed line: least-squares fit.}
\Description{A scatter plot with Video-MME overall accuracy on the horizontal axis, ranging from about 62 to 89, and mean AD-rating proficiency theta on the vertical axis, ranging from about negative 1.45 to positive 0.9. Seven vision-language models are plotted. GPT-4o mini, shown in green, sits at the upper left with the lowest benchmark score of 64.8 but the highest rating proficiency of positive 0.55. Four models cluster in the middle with benchmark scores between 70.6 and 75.0 and proficiency values near zero: LLaVA-Video-72B at negative 0.08, GPT-4o at positive 0.21, Qwen2.5-VL at positive 0.16, and Gemini 1.5 Pro at positive 0.09. Gemini 2.5 Pro and Gemini 2.5 Flash, shown in red, sit at the lower right with the highest benchmark scores of about 84.8 but the lowest proficiency values, negative 0.66 and negative 1.12 respectively. A dashed least-squares line slopes downward from upper left to lower right. The downward slope is driven largely by the two Gemini 2.5 points; the middle four vary little in proficiency across a five-point benchmark range.}
\label{fig:benchmark}
\end{figure}

The ability to evaluate AD quality appears distinct from general video understanding. Among the seven models in our cohort with reported Video-MME scores, benchmark performance and mean AD-rating proficiency showed a negative rank correlation (Spearman $\rho=-0.75$, $n=7$; Fig.~\ref{fig:benchmark}). Gemini 2.5 Pro and Flash achieved the highest Video-MME scores in this set, ranked near the bottom as AD raters, whereas GPT-4o mini, the lowest-scoring model on Video-MME, performed best as an AD rater. Given the small sample size, this correlation should be interpreted as illustrative rather than as a statistical test. We note that the Gemini 2.5 scores are developer-reported rather than obtained from the public leaderboard; excluding these models, the correlation decreases to $\rho=-0.40$, while the overall trend remains unchanged.

This divergence reflects the difference between the two tasks. Video-MME evaluates whether a model can answer questions about video content, whereas AD evaluation requires assessing another agent's description against a structured quality framework, including accessibility guidelines, relevance, timing, and descriptive choices. This reveals that selecting VLM evaluators based solely on general video benchmarks may be insufficient; specialized evaluation tasks require validating whether models can reliably apply domain-specific criteria.

We also observed substantial variation within the VLM group. The VLM ability distributions were broad across dimensions, indicating that models should not be treated as a homogeneous class of evaluators. Some dimensions showed clearer separation between higher- and lower-performing models, while others showed more continuous variation. Aggregating results across models can therefore obscure meaningful differences in evaluator capability. Future work should investigate whether these differences arise from model scale, training data, instruction tuning, or task-specific capabilities.

\subsection{Iterating on the Rating Framework}
Applying our six-dimension framework at scale revealed practical challenges. Informal comments from raters indicated that evaluating all six dimensions within a single task imposes a substantial cognitive burden, even for those with the necessary expertise. This reflects a tradeoff in the framework's design: comprehensiveness can come at the cost of consistent application. A framework that covers every facet identified by professional guidelines may be difficult for any individual rater, human or VLM, to apply uniformly, introducing variance that the workflow then has to absorb.

Moreover, while our six-dimension framework was grounded in professional AD guidelines, we need to examine the dimensions more closely to ensure they reflect what constitutes quality in practice. This is an iterative process. AD authoring has recently been described as a creative and interpretive practice \cite{walczak2017creative, schaefferlacroix2023beyond}, where subjectivity and stylistic nuance are part of what makes a description good rather than deviations from quality. Asking raters to deliver objective ratings on something that is itself inherently subjective, and to match the interpretive judgments of experts, sets the task up against itself. Rethinking the framework therefore means asking not only how to make it easier to apply, but whether each dimension can be rated consistently at all, and whether some should be restructured, split, or replaced.

These challenges underscore why this study serves as a necessary formative pilot. Before validating the framework with BLV end-users, we needed to stabilize its dimensions through iterative testing with accessibility experts and sighted raters. Accessibility research already draws disproportionately on BLV participants \cite{mack2021accessibility}, and imposing early-stage testing on them before the framework's baseline consistency is verified would be a poor use of that participation. This pilot allows us to bring a more refined framework to the BLV community. We plan to pilot streamlined versions of the framework with BLV end-users to identify which dimensions are both practical and meaningful to them. Rigorous psychometric validation (establishing reliability, criterion validity, and construct validity) will also require a substantially larger item pool than this study permitted. We are actively developing this next iteration in continued partnership with our blind consultants and accessibility experts.

\subsection{Designing Scalable Human-AI Evaluation}
Our study suggests three implications for the design of human-AI evaluation systems for AD quality:

\begin{itemize}
    \item Treat full-length AD quality as a multi-dimensional construct. Full-length AD introduces challenges that short-clip evaluation does not surface: consistency and prioritization across scenes, and especially formatting decisions (inline vs. extended narration, timing relative to dialogue) that only manifest at length. Evaluation tools must foreground these dimensions rather than collapsing quality into a single score or treating formatting as secondary to content.
    \item Use AI ratings to scale, not replace. In our study and given the current capabilities of the VLMs, these models can approximate expert judgments on certain dimensions, but their feedback lacks diagnostic value. This suggests a role for VLMs as first-pass filters, with humans providing targeted review.  
    \item Leverage hybrid evaluation strategies. With our VLM cohort and assessment framework, we observed that different raters (humans and VLMs) excel at different dimensions. While these patterns may shift with other models or rating frameworks, these findings suggest value in designing systems that combine evaluators with complementary strengths to improve coverage and reliability.  
\end{itemize}

Beyond audio description, the proposed workflow provides a general approach for evaluating AI systems in accessibility domains. Expert-labeled datasets are resource-intensive, and each new model must be evaluated against them again---a recurring burden in tasks such as image captioning and product identification for BLV users \cite{garg2026trained}. A potential use is item-targeted evaluation: because item difficulty and rater proficiency share a logit scale, IRT can identify which items are most informative about a rater's ability. Future models could then be evaluated on smaller, high-information subsets of previously annotated data, with expert ground truth as the reference standard. Importantly, the finding that matters is not that VLMs rate well or badly, but that the question is answerable: proficiency claims about AI raters can be tested against expert ground truth rather than assumed from benchmark standing or accepted from agreement statistics alone.

\section{Limitations}
Our study has three primary scope limitations. First, the item sample was modest: evaluating 30 audio descriptions (10 videos × 3 versions) across six dimensions required 180 ratings per respondent. This high cognitive load restricted our sample to 10 videos, which constrained our ability to examine fine-grained differences across genres and reduced the statistical power of our analysis. Second, while blind professionals shaped several aspects of this work during formative development, including the framework, corpus, and training materials, the main rating pool (25 humans, 19 VLMs) did not include BLV end-users, whose perspective remains the primary audience the workflow is ultimately designed to serve. Third, our IRT analysis revealed that some items distinguished respondent ability more effectively than others, with poorly fitting items introducing noise into the estimates. This variance suggests that certain dimensions may be interpreted inconsistently by raters or capture secondary constructs. Future work should investigate why some items misfit and whether excluding them before re-fitting the Rasch model improves measurement.

\section{Conclusion}

This paper presents a methodological workflow for assessing audio description quality in uninterrupted, full-length video. The workflow uses Item Response Theory to model rater ability and item difficulty against expert-established ground truth. Ratings follow a six-dimension framework that integrates formatting considerations informed by our accessibility experts and blind consultants. Our findings show that human and VLM raters achieve comparable alignment with expert ground truth, although human raters still perform slightly better. However, VLM-generated explanations lack the diagnostic depth of human feedback, which provides actionable, scene-specific insights. These results suggest the potential for hybrid evaluation systems in which VLMs, once shown to be reliable, perform scalable first-pass assessments while human reviewers provide the diagnostic feedback needed to improve description quality. More broadly, the proposed workflow provides a systematic way to evaluate and interpret AI ratings before incorporating them into quality control pipelines. As digital video continues to grow, such an approach could help accessibility evaluation keep pace with production demands while maintaining the standards required by blind and low-vision audiences.

\begin{acks}
The contents of this paper were developed under grants from Ability Central and the National Institute on Disability, Independent Living, and Rehabilitation Research (NIDILRR grant number 90REGE0018), with additional support from Northeastern University's Khoury College SP²ARK program (Silicon Valley PhD Pathway Apprenticeship for Research and Knowledge). We thank Joshua Miele, Charity Pitcher-Cooper, Yue-Ting Siu, and Tanja Milojevic for their contributions to this project.
\end{acks}

\bibliographystyle{ACM-Reference-Format}
\bibliography{references-new}
\appendix

\appendix
\section{AD Generation Prompt Templates}
\label{gen_prompts}
 
This section contains the full prompt templates used in our system, including the system prompt with audio description guidelines, scene-level generation, optimization, and extended description filtering prompts.

\subsection{System Prompt}
 
The system prompt establishes the audio describer role and injects domain-specific guidelines that govern all downstream generation steps.

\begin{lstlisting}[style=prompt, caption={Guidelines string.}, label={lst:guidelines}]
AUDIO DESCRIPTION GUIDELINES:
- Describe what you see in a concise, factual manner.
- Always read on-screen text exactly as it appears.
- Be factual, objective, and precise in your
  descriptions.
- Use proper terminology and names from the context
  when possible.
- Match the tone and mood of the video.
- Do not over-describe -- less is more.
- Do not interpret or editorialize about what you see.
- Do not give away surprises before they happen.

CHARACTER IDENTIFICATION:
- When you recognize a character from the context,
  ALWAYS use their specific name.
- Before each scene, carefully review context to
  identify all named characters.
- Use the most specific identification possible based
  on the context information.
\end{lstlisting}

\begin{lstlisting}[style=prompt, caption={System message. The \texttt{\{guidelines\}} placeholder is replaced with the string in Listing~\ref{lst:guidelines}.}, label={lst:system-prompt}]
You are an expert audio describer creating descriptions for videos to enhance accessibility for blind and low-vision users. Your task is to generate video descriptions that are true, contextually meaningful, useful, well-timed, and polished. You must follow the audio description guidelines below.

{guidelines}
\end{lstlisting}

\subsection{Scene-Level Generation Prompt}
 
This prompt is issued once per scene. It receives the scene duration and narrative context, then requests structured JSON output for both on-screen text events and visual description events.

\begin{lstlisting}[style=prompt, caption={Prompt for generating scene-level descriptions and text events}]
SCENE DURATION: {scene_duration:.2f} seconds
 
CONTEXT:
{context}
 
You are analyzing a video scene. Identify specific
characters, locations, and any important elements
mentioned in the context.
 
First, generate a JSON array of Text on Screen events
("type": "Text on Screen"):
- Capture ALL visible on-screen text.
- DO NOT include transcript or dialogue.
- For each text event, include the EXACT start_time
  in seconds when the text appears.
 
INCLUDE: Titles, headings, names; informational text;
important dates or events.
EXCLUDE: Brand logos and watermarks; network logos;
social media handles; copyright notices.
 
Second, generate a JSON array of Visual events:
- Provide contextually rich visual description of the
  scene using concise wording.
- Describe each action in detail.
- ALWAYS use specific character names from context
  (not "person" or "woman").
- Focus on key actions, settings, and objects not
  mentioned in a previous description.
- Include clear start times for each visual event.
- DO NOT describe Text on Screen.
 
OUTPUT FORMAT (JSON array):
  - start_time  (in seconds)
  - type        ("Text on Screen" or "Visual")
  - text        (description or on-screen text)
\end{lstlisting}

\subsection{Inline Optimization Prompt}
 
When the estimated speech duration of a generated description exceeds the available silence gap, the following prompt asks the model to condense the text to fit the timing constraint.

\begin{lstlisting}[style=prompt, caption={Prompt for condensing descriptions to fit available speech time.}]

You are optimizing a set of visual descriptions for
a video.
 
ORIGINAL DESCRIPTIONS: "{combined_text}"
AVAILABLE TIME: {available_duration:.2f} seconds
 
TASK:
Combine and condense these descriptions to fit within
{available_duration:.2f} seconds of spoken audio.
 
GUIDELINES:
- Create a coherent, flowing description.
- Maintain the action order and use concise but
  natural language.
- Ensure the final description can be spoken within
  the time limit.
 
OUTPUT: Provide only the optimized description text,
without explanations.
\end{lstlisting}

If the first optimization attempt still exceeds the time budget (verified via TTS), this follow-up prompt provides the measured overshoot and requests further reduction.

\begin{lstlisting}[style=prompt, caption={Retry prompt for further reduction if the first attempt exceeds the time budget.}]
You are optimizing visual descriptions for a video.
 
PREVIOUS ATTEMPT: "{optimized_text}"
 
This description takes {tts_duration:.2f} seconds to
speak, but only {available_duration:.2f} seconds are
available. Reduce by {tts_duration -
available_duration:.2f} seconds.
 
TASK:
Create a SHORTER version that fits within the time
limit.
 
GUIDELINES:
- Keep the most critical visual elements.
- Eliminate redundant details.
- Use concise but natural language.
 
OUTPUT: Provide only the shortened description,
nothing else.
\end{lstlisting}

\subsection{Extended Description Filtering Prompt}
 
For scenes where inline timing is insufficient, this prompt evaluates candidate descriptions against the transcript and cumulative context to decide whether an extended (pause-the-video) description is warranted.

\begin{lstlisting}[style=prompt, caption={Prompt for filtering extended audio description candidates.}]
You are an accessibility expert selecting ONE visual
description per scene to convert to audio description
for blind and low-vision users.
 
CONTEXT:
- AD should be minimal and only interrupt audio when
  necessary.
- Spoken transcript is the primary information source.
- Use AD only when critical visual information is
  missing from the audio.
 
INPUT:
CURRENT SCENE TRANSCRIPT: {transcript_text}
CUMULATIVE TRANSCRIPT:    {cumulative_transcript}
CUMULATIVE DESCRIPTION:   {previous_desc_text}
VISUAL DESCRIPTIONS TO EVALUATE: {clip['text']}
 
EVALUATION CRITERIA (mark necessary = true if any):
- Conveys visual details absent from the audio.
- Describes important silent actions or key events.
- Identifies characters or locations that are
  otherwise ambiguous.
- Introduces a novel or distinct visual element.
- Notes scene changes or unannounced time jumps.
 
OUTPUT FORMAT (JSON array, one item per description):
  - id         (index of the description)
  - necessary  (true or false)
  - reason     (short explanation)
\end{lstlisting}

\section{Multi-dimensional Rating Framework for Audio Description}
\label{Appendix:framework}
This appendix section reproduces the complete documentation provided to expert raters.

\textbf{Overview}: 
This framework evaluates the effectiveness and enjoyment of both human-authored and AI-generated audio description (AD) for YouTube videos across two main dimensions: \textbf{Content} (four criteria) and \textbf{Formatting} (two criteria).

\subsection*{Rating Terminology}

\begin{itemize}
    \item[5:] \textbf{Just Right} — The description was excellent and seamless; no issues were detected.
    \item[4:] \textbf{Minor Issue} — Error was infrequent and barely noticeable, not meaningfully affecting the overall experience.
    \item[3:] \textbf{Perceptible Issue} — Error was noticeable and affected quality enough to reduce enjoyment, though the content was still understandable overall.
    \item[2:] \textbf{Major Issue} — Error was noticeable and significantly affected quality, requiring full attention to overcome.
    \item[1:] \textbf{Critical Issue} — Error severely broke the experience, making the content confusing or impossible to follow.
\end{itemize}

\subsection*{I. Content}
This section evaluates how well the descriptions explain what is happening.

\textit{(1) Accurate — Error-Free Content}

\textbf{Definition:} Description provides error-free visual information with correct identification of what's actually happening. No factual mistakes or misleading information.

\textbf{Listener Experience Question:} Were the descriptions factually correct and easy to understand? (i.e., Were character names, locations, and actions identified correctly without jargon or mistakes?)

\textbf{Evaluation Criteria (1–5):}  
\begin{itemize}
    \item[5:] Just Right — All visual elements seemed correct; no clear mistakes.
    \item[4:] Minor Issue — Mostly correct, with a few confusing parts.
    \item[3:] Perceptible Issue — Generally correct but with noticeable errors.
    \item[2:] Major Issue — Several factual errors or confusing parts.
    \item[1:] Critical Issue — Major errors or misleading info throughout.
\end{itemize}

\textit{(2) Prioritized — Context \& Inference}

\textbf{Definition:} The description achieves optimal prioritization by selecting details based on their contextual significance and inferential value. For example, the description prioritizes contextually-rich details over generic descriptions such as ``Harry runs into the forest'' vs. ``a boy runs into the forest'', and makes reasonable inferences, such as ``a boy in a soccer uniform'' vs. ``a boy in a red jersey and tall socks''.

\textbf{Listener Experience Question:} Did the descriptions tell you the most important things you needed to know, without too much extra detail? (i.e., Did the description leave out essential action, create long pauses where visuals weren't described, OR include so many vague or unnecessary details that it became distracting?)

\textbf{Evaluation Criteria (1–5):}  
\begin{itemize}
    \item[5:] Just Right — Told you exactly what you needed to know to follow along, with no wasted words.
    \item[4:] Minor Issue — Missing some very small details OR slightly verbose/vague in a few spots.
    \item[3:] Perceptible Issue — Noticeable imbalance: some visual details were left undescribed during pauses, or extra/unnecessary information reduced clarity.
    \item[2:] Major Issue — Missing some important details OR included too many unneeded, vague details.
    \item[1:] Critical Issue — Major gaps in essential information OR excessively verbose, distracting detail throughout.
\end{itemize}

\textit{(3) Consistent — Consistency \& Coherence}

\textbf{Definition:} The description maintains consistent terminology, style, and tone, supporting a coherent and unified narrative throughout the video.

\textbf{Listener Experience Question:} Did the description maintain a consistent style, tone, and use of terms? (i.e., Was the language level right for the content, and did the describer use the same name/term for the same person/object throughout?)

\textbf{Evaluation Criteria (1–5):}  
\begin{itemize}
    \item[5:] Just Right — Tone and terminology remained smooth and fully consistent.
    \item[4:] Minor Issue — Mostly consistent, with rare, small variations in tone/style.
    \item[3:] Perceptible Issue — Adequate consistency, but some noticeable shifts in style or minor inconsistency in terminology.
    \item[2:] Major Issue — Frequent inconsistencies in word choice or tone, making the narrative difficult to follow.
    \item[1:] Critical Issue — Wildly inconsistent tone/style, or key terms/names shifted, making the narrative disjointed.
\end{itemize}

\textit{(4) Equal — Objectivity \& Non-Interpretation}

\textbf{Definition:} The description ensures equal access by being objective and without personal interpretation, bias, or unnecessary commentary.

\textbf{Listener Experience Question:} Did the description stick to objective facts and actions, without adding personal bias or unnecessary interpretation? (i.e., Did the describer tell you what to think instead of what they saw?)

\textbf{Evaluation Criteria (1–5):}  
\begin{itemize}
    \item[5:] Just Right — Completely objective; focused only on the visible facts.
    \item[4:] Minor Issue — Generally objective with rare, minor subjective/interpretive moments.
    \item[3:] Perceptible Issue — Interpretive or subjective language was evident enough to affect neutrality, but not so frequent that it overtook the description.
    \item[2:] Major Issue — Some subjective or unnecessary commentary was present.
    \item[1:] Critical Issue — Highly interpretive or biased language interfered significantly with the story.
\end{itemize}

\subsection*{II. Formatting}
This section evaluates the technical delivery and synchronization of descriptions.

\textit{(1) Strategic Use of Description Method (Inline vs. Extended)}

\textbf{Definition:} The description makes effective choices between inline and extended description methods based on content characteristics.

\textbf{Inline} description is the standard and preferred method when:
\begin{itemize}
    \item Sufficient natural pauses exist within original content timing \cite{NationalCenterforAccessibleMedia2017AccessibleGuidelines}
    \item Visual content can be adequately described within available audio gaps \cite{3PlayMedia2020AudioGuidelines}
\end{itemize}

\textbf{Extended} description is appropriate when:
\begin{itemize}
    \item Text-heavy videos, like recordings of slideshows or lectures \cite{3PlayMedia2020AudioGuidelines}
    \item Dialogue-heavy videos, as audio description shouldn't drown out what people are saying \cite{3PlayMedia2020AudioGuidelines}
    \item Noisy videos containing important music or sound, as audio description could detract from these elements \cite{3PlayMedia2020AudioGuidelines}
    \item Videos with short cuts and/or extremely detailed frames where standard description would be incomplete by the next cut \cite{3PlayMedia2020AudioGuidelines}
    \item Essential visual information cannot adequately fit within available natural pauses: ``If no such pause exists, you must insert an extended description at that point'' \cite{NationalCenterforAccessibleMedia2017AccessibleGuidelines}
\end{itemize}

\textbf{Listener Experience Question:} Was the extended description used appropriately for text, complex scenes, or when time was otherwise tight?

\textbf{Evaluation Criteria (1–5):}  
\begin{itemize}
    \item[5:] Just Right — The decision to pause (or not to pause) was perfectly done, ensuring all information was delivered well.
    \item[4:] Minor Issue — Pauses were mostly correct but felt slightly misplaced or unneeded a few times.
    \item[3:] Perceptible Issue — Pauses included some poor choices, either sometimes using extended AD unnecessarily or missing opportunities when an extended description was needed.
    \item[2:] Major Issue — Pauses were used unnecessarily or key information was rushed because a pause was not used when needed, which affected the understanding of the contents.
    \item[1:] Critical Issue — Frequent, unnecessary pauses OR key information was consistently rushed because an extended pause was not used when required.
\end{itemize}

\textit{(2) Timing \& Placement}

\textbf{Definition:} Appropriate timing of description placement relative to visual content and audio elements based on established accessibility standards. Timing standards for both description methods as follows:
\begin{itemize}
    \item No interruption of important dialogue or essential sound effect \cite{2024DescribedDCMP}.
    \item Insert descriptions at natural points in the timeline — don't cut off speakers mid-word, but take advantage of brief pauses. Even pauses between words or sentences suffice as long as the description is not out of context \cite{NationalCenterforAccessibleMedia2017AccessibleGuidelines}.
    \item Place descriptions as close to the visual action as possible \cite{3PlayMedia2020AudioGuidelines}.
    \item Pre-description is allowed: descriptions may be inserted ``slightly before the action occurs on screen'' if it clarifies the situation \cite{NationalCenterforAccessibleMedia2017AccessibleGuidelines}.
\end{itemize}

\textbf{Listener Experience Question:} Were the descriptions placed seamlessly, without cutting off dialogue or sound effects? (i.e., Were the descriptions too early or too late relative to the action, or did they interrupt the original audio?)

\textbf{Evaluation Criteria (1–5):}  
\begin{itemize}
    \item[5:] Just Right — Optimal timing; seamless and never interrupted the original audio.
    \item[4:] Minor Issue — Generally good placement, but descriptions were occasionally too early/late or slightly overlapped non-essential audio.
    \item[3:] Perceptible Issue — Mistimed delivery was clear and occasionally interfered with audio, reducing the seamlessness of the experience.
    \item[2:] Major Issue — Descriptions often mistimed or occasionally overlapped with important dialogue/sound effects.
    \item[1:] Critical Issue — Descriptions frequently interrupted dialogue or were consistently too early/late for the action.
\end{itemize}

\section{VLM Respondents Prompt}
\label{appendix:prompts}

\begin{lstlisting}[style=prompt, caption={Prompt used to instruct VLMs to apply the six-dimension framework for evaluating audio description quality.}]

PROMPT_FOR_EVALUATION = """
ROLE: You are an expert Accessibility Consultant specializing in evaluating the listener experience of audio description (AD) for video content.

CONTEXT: I am providing you with two assets:
1. A video file.
2. The structured JSON data of the existing audio description, which is included below.

**JSON DATA:**
```json
{json_data}
```

TASK: Analyze the video and the JSON data to evaluate the quality of the audio description track using the Multi-Dimensional Rating Framework for Audio Description.

RATING FRAMEWORK:
This framework evaluates audio description across two main dimensions:
I. CONTENT (4 criteria based on DCMP guidelines)
II. FORMATTING (2 criteria covering how and when descriptions are delivered)

Scale (1 - 5 ):
1 = Critical Issue, 2 = Major Issue, 3 = Perceptible Issue, 4 = Minor Issue, 5 = Just Right

I. CONTENT CRITERIA:

1. ACCURATE - Error Free Content
Definition: Description provides error-free visual information with correct identification of what's actually happening. No factual mistakes or misleading information.

5: All visual elements are factually correct. No errors in describing what's actually happening. Perfect factual accuracy.
4: Mostly factually correct with minor errors that don't mislead. Generally accurate descriptions.
3: Generally factually correct but with some noticeable errors. Mostly accurate with some mistakes.
2: Multiple factual errors that mislead about what's happening. Poor accuracy in descriptions.
1: Major factual errors or completely incorrect information. Fails to accurately describe what's happening.

2. PRIORITIZED - Context & Inference
Definition: The description achieves optimal prioritization by selecting details based on their contextual significance and inferential value. Prioritizes contextually-rich details over generic descriptions and makes reasonable inferences.

5: Just right balance - perfect prioritization on most significant elements for understanding. Chooses contextually relevant details and appropriate spatial information.
4: Good prioritization but not perfect - either slightly too generic or slightly excessive. Generally good choices about what to include.
3: Adequate prioritization but noticeable imbalance - either missing some important details or including some unnecessary information.
2: Poor prioritization - either incomplete important information or includes too many unimportant details. Poor choices about what matters.
1: Major problems - either major gaps in important information or describes everything including unimportant elements. No clear prioritization on what's significant.

3. CONSISTENT - Consistency & Coherence
Definition: The description maintains consistent terminology, style, and tone, supporting a coherent and unified narrative throughout the video.

5: Fully consistent in terminology and style. Narrative flows smoothly and coherently.
4: Mostly consistent with minor variations. The narrative remains generally coherent.
3: Adequate consistency, but some noticeable shifts in terminology or style.
2: Frequent inconsistencies in word choice or tone. The narrative becomes difficult to follow.
1: No consistency maintained. The narrative is disjointed or incoherent.

4. EQUAL - Objectivity & Non-Interpretation
Definition: The description ensures equal access by being objective and without personal interpretation, bias, or unnecessary commentary.

5: Completely objective. No personal interpretation. Appropriate descriptive language without editorial comment.
4: Generally objective with rare minor interpretive moments.
3: Mostly objective but some unnecessary interpretation present.
2: Frequent interpretive language. Some bias evident in descriptions.
1: Highly interpretive and biased. Significant personal commentary interferes with equal access.

II. FORMATTING CRITERIA:

1. Strategic Use of Description Method (Inline vs. Extended)
Definition: The description makes effective choices between inline and extended description methods based on content characteristics.
Notes:
- Inline description is preferred when sufficient natural pauses exist and visual content can be adequately described within available audio gaps
- Extended description is appropriate for text-heavy videos, dialogue-heavy content, noisy videos with important music/sound, videos with short cuts/detailed frames, or when essential visual information cannot fit within natural pauses

5: Perfect method selection - consistently chooses inline for content with adequate pauses, extended only when absolutely necessary based on professional criteria
4: Good method selection with occasional minor errors - generally appropriate choices with rare unnecessary use of extended description
3: Adequate method selection but some poor choices - sometimes uses extended unnecessarily or misses opportunities when extended is needed
2: Poor method selection - frequently uses wrong method, either overusing extended description or failing to use it when required
1: Severe method selection issues - no understanding of when to use inline vs. extended based on professional standards

2. Timing & Placement
Definition: Appropriate timing of description placement relative to visual content and audio elements based on established accessibility standards.
Notes:
- No interruption of important dialogue or essential sound effects
- Insert descriptions at natural points in the timeline
- Place descriptions as close to the visual action as possible
- Pre-description is allowed if it clarifies the situation

5: Optimal timing - descriptions placed during natural pauses close to the visual action without interrupting essential audio
4: Occasionally poor timing - generally good placement but sometimes descriptions are too early, too late, or slightly overlap important audio
3: Noticeable timing issues - descriptions poorly timed relative to visual content, some interference with dialogue
2: Poor timing - descriptions often mistimed, frequently interrupting dialogue or placed too far from relevant action
1: Severe timing issues - consistently poor timing that disrupts content flow and interferes with essential audio

OUTPUT FORMAT:
You MUST return your response as a single, flat JSON object. Do not use nested structures. Do not include any text or markdown before or after the JSON. The JSON object must have this EXACT structure:
{{
"accurate_rating": "1-5",
"accurate_justification": "Justification for the rating.",

"prioritized_rating": "1-5",
"prioritized_justification": "Justification for the rating.",

"consistent_rating": "1-5",
"consistent_justification": "Justification for the rating.",

"equal_rating": "1-5",
"equal_justification": "Justification for the rating.",

"strategic_method_selection_rating": "1-5",
"strategic_method_selection_justification": "Justification for the rating.",

"timing_and_placement_rating": "1-5",
"timing_and_placement_justification": "Justification for the rating."
}}
"""
\end{lstlisting}

\clearpage

\begin{table*}[t]
\section{Person-level Fit Statistics}
\vspace{1em}
\label{Appendix:person_fit_stats}
\centering
\small
\renewcommand{\arraystretch}{0.8}
\begin{tabular}{lcccccc}
\toprule
\textbf{Respondent} & \textbf{Accurate} & \textbf{Prioritized} & \textbf{Consistent} & \textbf{Equal} & \textbf{Description Method} & \textbf{Timing \& Placement} \\
\midrule

\multicolumn{7}{l}{\textit{Human Respondents}} \\
1 & 0.72058 & 0.85944 & 0.73279 & 1.02785 & 0.72584 & 0.74522 \\
2 & 0.77284 & 0.89461 & 0.50876 & 0.72790 & 0.83733 & 0.85684 \\
3 & 0.90876 & 1.32433 & 0.78353 & 1.22334 & \textbf{1.42988} & 1.08812 \\
4 & 0.78827 & 0.85915 & 0.73359 & 0.69606 & 0.61433 & 0.69368 \\
5 & 0.81951 & 0.56413 & 0.79678 & 0.90369 & 0.42937 & 1.00184 \\
6 & 0.74901 & 0.95671 & 0.58189 & 0.85290 & 0.68529 & 1.01874 \\
7 & 0.67985 & 0.71703 & 0.71613 & 0.92510 & 0.50081 & 0.77330 \\
8 & 0.65292 & 0.68830 & 0.66860 & 1.32618 & 0.61629 & 0.89211 \\
9 & 0.63659 & 0.95406 & 0.98180 & 1.13994 & 1.23492 & 1.20499 \\
10 & 0.53418 & 0.74861 & 0.85145 & 0.64924 & 0.75879 & 0.89813 \\
11 & 0.64536 & 0.69474 & 0.79494 & 0.90556 & 0.75474 & 0.71871 \\
12 & 0.77208 & 0.66269 & 0.69907 & 0.59429 & 0.53209 & 0.71188 \\
13 & 0.79580 & 0.69919 & 0.69748 & 0.64636 & 0.64862 & 0.77071 \\
14 & 0.89370 & 0.91225 & 0.76059 & 0.87399 & 0.87842 & 0.74555 \\
15 & 0.49008 & 0.67881 & 0.82784 & 0.74125 & 0.77592 & 0.81032 \\
16 & 1.06307 & 0.89341 & 0.63230 & 0.76364 & 0.74565 & 0.67959 \\
17 & 0.59434 & 1.08866 & 0.77292 & 1.02668 & \textbf{1.36527} & 0.81321 \\
18 & 0.72727 & 0.56106 & 0.72003 & 0.78556 & 0.82073 & 0.80591 \\
19 & 0.62883 & 0.79842 & 0.60069 & 0.66556 & 0.83109 & 0.61800 \\
20 & 0.85953 & 0.60744 & 0.93812 & 1.03458 & 0.89485 & 0.73318 \\
21 & 0.77226 & 0.86467 & 0.64102 & 0.73100 & 0.99929 & 0.64115 \\
22 & 1.11690 & 0.89898 & 1.12275 & 1.09441 & 0.82076 & \textbf{1.41764} \\
23 & 0.71853 & 0.90829 & 0.75332 & 0.94902 & \textbf{1.35911} & 0.89200 \\
24 & 0.68245 & 0.89155 & 0.67234 & 0.64683 & 1.05510 & 1.08885 \\
25 & \textbf{1.85245} & 1.27301 & 1.30313 & \textbf{2.56132} & 1.09665 & \textbf{1.44250} \\

\midrule
\multicolumn{7}{l}{\textit{VLM Respondents}} \\

claude-opus-4-6 & \textbf{1.65617} & 1.09371 & 1.26856 & 1.23946 & 0.81760 & 1.31978 \\
claude-sonnet-4-6 & \textbf{2.11322} & \textbf{4.88754} & \textbf{2.18471} & \textbf{2.10660} & \textbf{1.42006} & \textbf{3.11941} \\
gemini\_1.5\_pro & 1.30642 & 1.25772 & 1.01805 & 0.88947 & 1.07259 & 0.94454 \\
gemini\_2.5\_flash & \textbf{1.60429} & \textbf{1.46701} & \textbf{1.34969} & 1.07388 & \textbf{1.33210} & \textbf{2.99294} \\
gemini\_2.5\_pro & 1.09793 & 1.17624 & 1.18692 & 0.62552 & 1.32923 & \textbf{1.35975} \\
gemini\_3.0\_flash & 0.67964 & 1.24517 & 0.82033 & 0.52818 & 1.24344 & 0.58326 \\
gemini\_3.0\_pro & \textbf{1.40076} & \textbf{1.59977} & 0.93868 & 0.72596 & \textbf{1.34433} & \textbf{3.88454} \\
gpt4o & 0.87060 & 0.81863 & 1.05390 & 1.03766 & 0.75687 & 0.69440 \\
gpt4o\_mini & 0.59813 & 0.82536 & 0.69169 & 0.63847 & 0.63451 & 0.83031 \\
gpt5.2 & 1.17344 & \textbf{1.63503} & \textbf{1.69875} & 0.82156 & 1.09884 & \textbf{1.74796} \\
gpt5\_mini & 0.99500 & 0.72943 & 0.99036 & 0.88058 & 1.16051 & \textbf{1.41576} \\
gpt5 & 1.12463 & 1.19507 & \textbf{1.49164} & 1.23720 & 0.57573 & \textbf{1.56655} \\
llama3.2 & 0.86020 & 0.92093 & 0.91445 & 1.14753 & 0.54563 & 0.98869 \\
phi4 & 1.19590 & 1.03393 & \textbf{3.83058} & \textbf{3.99004} & \textbf{1.72593} & 0.73005 \\
qwen\_2.5 & 0.84134 & 0.81092 & 0.74514 & 0.73885 & 1.26248 & 0.66695 \\
qwen3 & 0.73847 & 1.07184 & 0.80973 & 0.88971 & 0.97422 & 0.71598 \\
NVILA-15B & 0.65381 & 1.07579 & 0.95095 & 0.49396 & 1.13963 & 0.85902 \\
llava-video-72B & 0.83280 & 0.94525 & 0.82789 & 0.96128 & 0.98003 & 0.87935 \\
InternVL2\_5-26B & 0.72964 & 1.03337 & 0.84469 & \textbf{1.80469} & \textbf{1.60632} & 0.67513 \\

\bottomrule
\end{tabular}%
\caption{Fit statistics (MNSQ) across six evaluation dimensions. Values $\geq 1.33$ are highlighted in bold, indicating potential misfit.}
\Description{The Table represents the person-level fit statistics (Mean-Square Fit Statistics) of 44 respondents including humans and VLMs on 6 dimensions. Each respondent has the value on each dimension and the values above 1.33 indicates potential misfit where the respondent's pattern is random. All the values above 1.33 is bolded.}
\label{tab:fit_statistics}
\end{table*}

\clearpage

\begin{table*}[h]
\section{Item-level Fit Statistics}
\vspace{1em}
\label{Appendix:item-fit}
\small
\renewcommand{\arraystretch}{0.8}
\setlength{\tabcolsep}{4pt}
\begin{tabular}{c cccccc}
\toprule
\textbf{Item} & \textbf{Accurate} & \textbf{Prioritized} & \textbf{Consistent} & \textbf{Equal} & \textbf{Description Method} & \textbf{Timing \& Placement} \\
\midrule

1  & 0.97 & 0.91 & 0.73 & 0.81 & 1.26 & 0.98 \\
2  & 0.82 & 1.13 & 0.84 & 0.82 & 1.32 & 0.67 \\
3  & 0.94 & 0.96 & 0.94 & 0.85 & 0.88 & 0.78 \\
4  & 1.03 & 1.03 & \textbf{1.34} & 0.85 & 1.18 & \textbf{1.42} \\
5  & 1.00 & 1.21 & 0.95 & 0.76 & 0.75 & 0.88 \\
6  & 1.05 & 0.91 & 1.12 & 0.75 & 0.90 & 0.85 \\
7  & 1.17 & \textbf{1.49} & 0.93 & \textbf{1.86} & 1.23 & 0.91 \\
8  & 0.86 & 0.80 & 0.91 & 1.07 & 0.79 & 1.03 \\
9  & 1.14 & 0.91 & 0.86 & \textbf{1.69} & 0.81 & 0.66 \\
10 & 1.26 & 0.90 & 0.97 & 0.87 & 0.97 & 1.26 \\
11 & 0.88 & 0.85 & 0.87 & 1.03 & 1.26 & 0.90 \\
12 & 1.12 & 0.92 & 0.93 & 0.94 & 0.77 & 0.86 \\
13 & 1.20 & 0.93 & 0.87 & 0.84 & 0.99 & 0.98 \\
14 & 1.03 & 1.22 & 0.89 & 0.85 & 1.12 & 0.81 \\
15 & 0.89 & 1.05 & 1.07 & 0.80 & 1.22 & 0.95 \\
16 & 1.08 & \textbf{1.49} & 1.31 & 1.20 & 1.03 & \textbf{1.79} \\
17 & 0.98 & 1.06 & 0.92 & 0.99 & 0.91 & 1.14 \\
18 & 1.08 & 0.91 & 0.69 & 0.98 & 0.87 & 0.99 \\
19 & 0.96 & 1.01 & 1.11 & 1.14 & 1.02 & 1.25 \\
20 & \textbf{1.36} & 1.14 & 1.05 & 1.10 & 1.02 & 1.09 \\
21 & 0.78 & 0.82 & 0.88 & 0.99 & 1.09 & 0.90 \\
22 & 0.77 & 0.85 & 1.07 & 0.74 & 0.80 & 0.92 \\
23 & 0.84 & 0.89 & 0.74 & 1.20 & 0.84 & 0.77 \\
24 & 1.01 & 0.93 & 0.86 & 0.76 & 0.88 & 0.84 \\
25 & 1.03 & 0.85 & 0.94 & 0.78 & 1.16 & 1.19 \\
26 & 0.84 & 0.87 & \textbf{1.54} & 0.87 & 0.78 & 1.01 \\
27 & 0.76 & 0.86 & 0.98 & 0.82 & 0.96 & 0.82 \\
28 & 0.89 & 0.90 & 0.81 & 1.32 & 1.10 & 0.89 \\
29 & 1.32 & 1.10 & 1.25 & 1.18 & 0.99 & \textbf{1.37} \\
30 & 0.82 & 0.86 & 0.92 & 0.91 & 1.03 & 0.82 \\

\bottomrule
\end{tabular}
\caption{Item fit statistics (MNSQ) across six evaluation dimensions. Values $\geq 1.33$ are highlighted in bold, indicating potential misfit.}
\Description{The Table represents the item-level fit statistics (Mean-Square Fit Statistics) of the 30 items (ADs) on 6 dimensions. Each item has the value on each dimension and the values above 1.33 indicates potential misfit where the item's pattern of the difficulty is random. All the values above 1.33 is bolded.}
\label{tab:item_fit}
\end{table*}

\begin{table*}
\section{Proficiency Scores ($\theta$)}
\vspace{1em}
\label{Appendix:proficiency}
\centering
\small
\renewcommand{\arraystretch}{0.8}
\setlength{\tabcolsep}{4pt}
\begin{tabular}{lcccccc}
\toprule
\textbf{Respondent} & \textbf{Accurate} & \textbf{Prioritized} & \textbf{Consistent} & \textbf{Equal} & \textbf{Description Method} & \textbf{Timing \& Placement} \\
\midrule

\multicolumn{7}{l}{\textit{Human Respondents}} \\

1 & 0.997 & 0.773 & 0.446 & -0.025 & 0.858 & 0.857 \\
2 & -0.012 & 0.257 & 0.266 & -0.324 & 0.758 & 0.471 \\
3 & -0.535 & -0.363 & 0.180 & 0.182 & 0.157 & -0.289 \\
4 & 0.555 & 0.680 & 0.355 & -0.025 & 0.315 & 0.658 \\
5 & -0.188 & 0.099 & -0.070 & 0.632 & -0.143 & 1.074 \\
6 & -0.188 & 0.257 & 0.541 & -0.324 & 0.662 & 0.293 \\
7 & 0.658 & 0.680 & 0.446 & 0.289 & 0.570 & 0.658 \\
8 & 0.454 & 0.178 & -0.151 & 1.684 & 0.482 & 0.857 \\
9 & 1.122 & 0.257 & 0.266 & 0.400 & 0.235 & 0.038 \\
10 & 0.357 & 0.338 & 0.180 & 1.022 & 0.482 & 0.381 \\
11 & -0.101 & 0.505 & 0.355 & -0.709 & 0.482 & 0.293 \\
12 & 0.357 & -0.209 & 0.012 & -0.225 & 0.235 & 0.122 \\
13 & 0.262 & 0.421 & 0.355 & -0.225 & 0.315 & 0.293 \\
14 & 1.256 & 0.969 & 0.541 & 0.077 & 1.327 & 1.601 \\
15 & 0.394 & 0.850 & 1.101 & -0.047 & -0.040 & 0.070 \\
16 & -0.012 & 0.421 & 0.541 & -0.421 & 0.758 & 0.658 \\
17 & -0.073 & -0.687 & 0.075 & 0.377 & -0.083 & -0.390 \\
18 & 0.658 & 0.257 & 0.541 & -0.421 & 0.157 & 0.857 \\
19 & 1.357 & 0.428 & 0.319 & -0.742 & 0.857 & 0.655 \\
20 & 0.262 & -0.133 & 0.095 & -0.025 & 0.397 & 0.122 \\
21 & 0.078 & 0.421 & -0.070 & -0.324 & 0.397 & -0.045 \\
22 & -0.012 & -0.440 & -0.151 & 0.756 & -0.215 & 0.207 \\
23 & 0.658 & 0.680 & 0.541 & -0.225 & 0.235 & 0.381 \\
24 & -0.012 & -0.834 & 0.095 & 1.022 & -0.430 & -0.045 \\
25 & -1.359 & -0.517 & -0.707 & -1.802 & -0.939 & -0.612 \\

\midrule
\multicolumn{7}{l}{\textit{VLM Respondents}} \\

claude-opus-4-6 & -0.887 & -0.286 & -0.469 & -0.421 & -0.864 & -0.370 \\
claude-sonnet-4-6 & -1.164 & -2.067 & -1.391 & -2.020 & -1.510 & -1.374 \\
gemini\_1.5\_pro & 0.357 & -0.133 & -0.469 & 0.886 & -0.287 & 0.207 \\
gemini\_2.5\_flash & -1.164 & -0.674 & -0.151 & 0.289 & -2.129 & -2.911 \\
gemini\_2.5\_pro & -1.070 & -0.595 & 0.095 & 0.886 & -1.092 & -2.200 \\
gemini\_3.0\_flash & -0.710 & -0.363 & 0.012 & 0.886 & 0.157 & -0.289 \\
gemini\_3.0\_pro & -0.798 & -0.363 & 0.266 & 0.756 & -0.864 & -1.969 \\
gpt4o & 0.169 & -0.363 & 0.446 & 0.756 & -0.215 & 0.471 \\
gpt4o\_mini & 0.555 & 0.099 & 0.446 & 1.168 & 0.157 & 0.857 \\
gpt5 & -0.710 & -0.056 & -0.787 & -0.805 & -0.790 & -0.370 \\
gpt5\_mini & 0.658 & -0.056 & -0.469 & 0.400 & -0.287 & -0.208 \\
gpt5.2 & -0.622 & -0.517 & -1.208 & -0.998 & -0.717 & -0.857 \\
InternVL2\_5-26B & -0.012 & 0.021 & -0.231 & -0.901 & 0.315 & -0.045 \\
llama3.2 & 0.154 & 0.107 & -0.031 & -0.172 & 0.002 & 0.200 \\
llava-video-72B & 0.357 & 0.178 & -0.151 & -0.324 & -0.260 & -0.289 \\
NVILA-15B & -0.535 & -0.363 & 0.180 & 1.022 & 0.005 & -0.531 \\
phi4 & -0.798 & -0.056 & -2.165 & -2.368 & -0.215 & -0.531 \\
qwen\_2.5 & 0.262 & 0.421 & 0.012 & -0.517 & 0.315 & 0.471 \\
qwen3 & -0.681 & -0.235 & -0.272 & 0.501 & 0.224 & 0.100 \\

\bottomrule
\end{tabular}
\caption{Respondent ability estimates ($\theta$) across six evaluation dimensions.}
\Description{The table represents the proficiency scores of 44 respondents including humans and VLMs on 6 dmensions. Each respondent has the value on each dimension and the higher values indicates the higher change or ability for that respondent to achieve the perfect agreement of that task. The values are relevent to each other only within the same dimension.}
\label{tab:theta_values}
\end{table*}    

\end{document}